\documentclass[12pt]{iopart}
\usepackage{amssymb}
\usepackage{graphicx}
\usepackage{subfigure}
 \usepackage{color}
\newcommand{\bra}[1]{\langle#1|}
\newcommand{\ket}[1]{|#1\rangle}

\providecommand{\openone}{\leavevmode\hbox{\small1\kern-3.8pt\normalsize1}}

\begin{document}

\title{Improving autonomous thermal entanglement generation using a common reservoir}

\author{Zhong-Xiao Man$^1$, Armin Tavakoli$^2$, Jonatan Bohr Brask$^2$ and Yun-Jie Xia$^1$}

\address{$^1$ School of Physics and Physical Engineering, Shandong Provincial Key Laboratory of Laser Polarization and Information Technology, Qufu Normal University, 273165, Qufu, China}
\address{$^2$ Department of Applied Physics, University of Geneva, 1211 Geneva, Switzerland}

\ead{zxman@qfnu.edu.cn}

\begin{abstract}
We study the entanglement generated in the steady state of two interacting qubits coupled to thermal reservoirs. We show that the amount of steady-state entanglement can be enhanced by the presence of a third thermal reservoir which is common to both qubits. Specifically, we find that entanglement can be enhanced as long as the temperature of the common reservoir is below the thermalisation temperature of the qubits, whenever a single temperature can be assigned to the steady state of the qubits in the absence of the common reservoir. Moreover, the amount of entanglement generated with the common reservoir present can be significantly larger than that which can be obtained without it for any temperature of the individual reservoirs. From the perspective of thermodynamics, we find that enhancement of entanglement is associated with heat absorption by the common reservoir. We propose a possible implementation of our scheme in superconducting circuits and find that a significant enhancement of steady-state entanglement should be observable under experimentally realistic conditions.
\end{abstract}

\section{Introduction}

Quantum entanglement is a fundamental concept in quantum mechanics as well as a key resource in quantum information science e.g.~for quantum communication, computation, and metrology \cite{ent,giovanetti2011}. Entanglement is notoriusly fragile in the presence of environmental noise, complicating the realisation of practical applications. Hence, understanding how to generate, protect, and enhance entanglement in different environments is important both fundamentally and for enabling quantum information technologies.

A large body of work has been devoted to enhancing and protecting entanglement via direct manipulation, for example through entanglement purification \cite{pur1,pur2,pur3}, quantum error correction\cite{error-corr1,error-corr2}, dynamical decoupling \cite{DD1,DD2,DD3}, or exploiting the quantum Zeno effect \cite{Zeno1,Zeno2}, or weak measurements \cite{weak1,weak2,weak3,weak4}. In addition to these strategies, which aim to counter the effects of noise, it turns out that dissipation can also be beneficial under certain conditions, and can be exploited for entanglement generation in both transient and steady regimes \cite{diss-pre1,diss-pre2,diss-pre3,diss-pre4,diss-pre5,diss-pre6} in various physical contexts \cite{phys-sys7,phys-sys1,phys-sys2,phys-sys3,phys-sys4,phys-sys5,phys-sys6}. Driven dissipative preparation of entangled states has been demonstrated experimentally for atomic ensembles \cite{atom}, trapped ions \cite{ion1,ion2}, and superconducting qubits \cite{super}.

Entanglement can also be generated thermally, without any driving. In a composite, interacting quantum system, the energetic ground state may be entangled, and hence cooling the system sufficiently will generate entanglement. In thermal equilibrium at higher temperatures, entanglement may still be present. In fact, the topic of how entanglement varies with temperature has long been a concern of condensed-matter physicists \cite{equi1,equi2,equi3,equi4,equi5,equi6,equi7}. In particular Refs.~\cite{equi1,equi2,equi3,equi4} studied the variation of entanglement with temperature and magnetic field in spin chains in thermal equilibrium.

Interestingly, entanglement can be enhanced by moving out of thermal equlibrium where temperature gradients induce energy currents among the interacting subsystems. Ref.~\cite{nonequi1} found increase in entanglement due to an energy current in a  spin chain. Ref.~\cite{nonequi2} studied changes in steady-state entanglement in a model of two interacting qubits coupled to different heat baths. The temperature gradient was shown to enhance or suppress entanglement depending on the internal coupling strength between the qubits. The dynamics of nonequilibrium thermal entanglement in a similar model was studied in Ref.~\cite{nonequi3} with particular attention to the case of non-resonant qubits, and Refs.~\cite{nonequi4,nonequi5} studied chains of three qubits out of equilibrium. Refs.~\cite{Brunner2014,BraskPRE2015,Brask2015,Tavakoli2017} demonstrated that entanglement can enhance the performance of quantum thermal machines, and that such machines can be harnessed for entanglement generation. In particular, in Ref.~\cite{Brask2015} a simple two-qubit thermal machine was presented which generates steady-state entanglement by operating between heat reservoirs at different temperatures. A similar two-qudit machine combined with filtering enables generation of maximal entanglement in any dimension when the temperature gradient is maximal \cite{Tavakoli2017}. All of these works confirm that there are strong connections between thermal entanglement and quantum thermodynamics.

While a lot can be learned from and achieved with coupled qubits in contact with independent heat reservoirs, in practical situations there will often be coupling to a common environment as well, and it is also interesting theoretically to understand the effects of such a shared reservoir. In fact, a common reservoir may itself enable entanglement generation. It was shown that entanglement between two qubits could be induced by a common, thermal, single-mode field \cite{MSKim}. Similarly, qubits in a common heat bath can become entangled when evolving through a purely noisy mechanism \cite{nonint1,nonint2}, and steady-state entanglement is found for qubits immersed in a common thermal reservoir \cite{equi7}. A common environment out of thermal equilibrium could lead to many-body entangled steady states \cite{BB1,BB2} and protect entanglement during evolution \cite{BB3}.

Here, we study thermal entanglement generation when both independent and common heat reservoirs are involved. We consider two interacting qubits coupled to individual heat reservoirs, as in the thermal machine of Ref.~\cite{Brask2015}, as well as to a common reservoir. We show that the steady-state entanglement can be enhanced by the presence of this common reservoir, and that the lower the temperature of the common reservoir, the larger the enhancement. The maximal critical temperature of the common reservoir enabling entanglement growth is the thermalized temperature of the coupled qubits if thermalization is achieved. Entanglement enhancement is accompanied by a thermodynamics process where heat is dissipated into the common reservoir. We also present a possible implementation of our scheme in superconducting circuits. We find that for experimentally accessible parameter settings, a significant improvement of steady-state entanglement can be realized.

\section{Model }

\begin{figure}[tbp]
\begin{center}
{\includegraphics[width=0.7\linewidth]{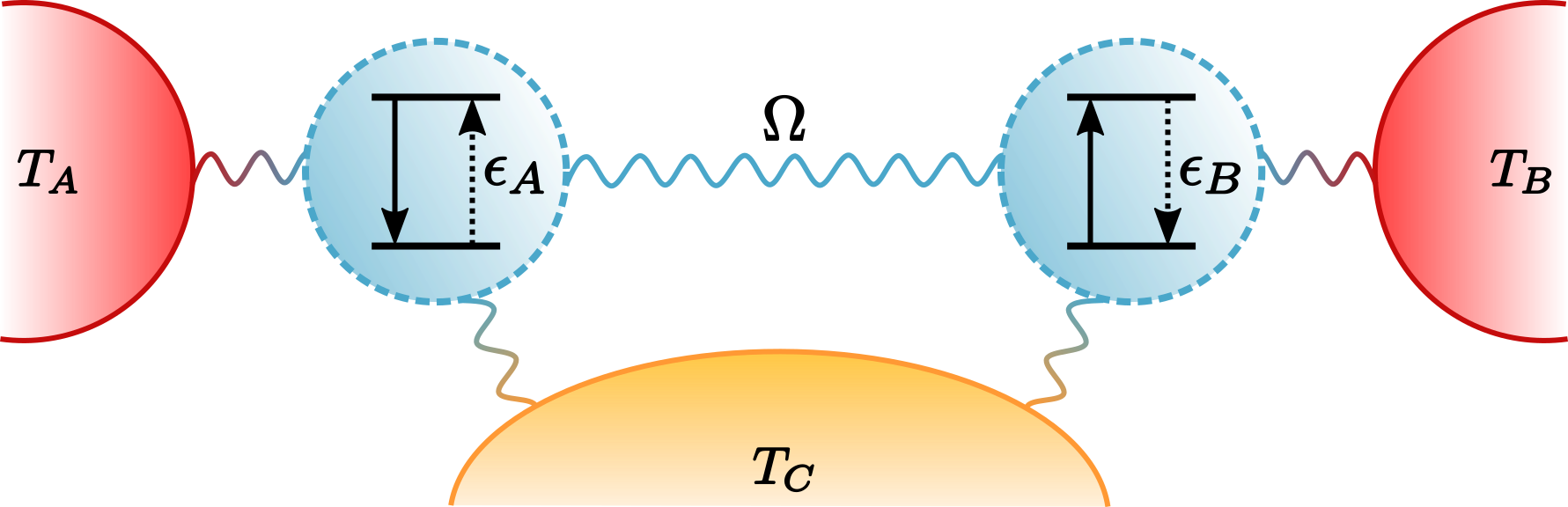} }
\end{center}
\caption{(Color online) Schematic diagram of the physical model under
consideration. Two qubits are coupled to each other with
a strength $\Omega$ and to two independent heat reservoirs with temperatures $T_{A}$
and $T_{B}$, respectively. A common reservoir with temperature $T_{C}$ is introduced to improve the
entanglement of the qubits.
}
\label{model}
\end{figure}

The system we consider, as depicted in Fig.\ref{model}, consists of two coupled qubits $A$ and $B$ interacting with two independent heat reservoirs $R_{A}$ and $R_{B}$, respectively, and potentially also to a common heat reservoir $R_{C}$. The Hamiltonian of the two qubits $ \hat{H}_{S}=\hat{H}_{0}+\hat{H}_{int} $ with the free Hamiltonian
\begin{equation}\label{H0}
\hat{H}_{0} = \epsilon_{A}\ket{1}_{A}\bra{1} \otimes \mathbb{I}_{B} + \epsilon_{B}\, \mathbb{I}_{A}\otimes \ket{1}_{B}\bra{1} ,
\end{equation}
and the interaction Hamiltonian
\begin{equation}\label{Hint}
\hat{H}_{int} = \Omega \left( \hat{\sigma}_{+}^{A}\otimes\hat{\sigma}_{-}^{B} + \hat{\sigma}_{-}^{A}\otimes\hat{\sigma}_{+}^{B} \right),
\end{equation}
where $\ket{0}_{\mu}$ and $\ket{1}_{\mu}$ are the ground and excited states of qubit $\mu \in\{A,B\}$ with energy gap $\epsilon_{\mu}$, $\mathbb{I}_{\mu}$ denotes the identity operator, $\hat{\sigma}_{+}^{\mu}=\ket{1}_{\mu}\bra{0}$ and $\hat{\sigma}_{-}^{\mu} = \ket{0}_{\mu}\bra{1}$ are the raising and lowering operators for qubit $\mu$, and $\Omega$ is qubit-qubit coupling strength.

The bosonic reservoirs are assumed to be thermal at temperatures $T_A$, $T_B$, and $T_C$. They are described by the Hamiltonian
\begin{equation}\label{H-reservoir}
\hat{H}_{R} = \sum_{l}\omega_{a,l}\hat{a}_{l}^{\dag}\hat{a}_{l} + \sum_{m}\omega_{b,m}\hat{b}_{m}^{\dag}\hat{b}_{m} + \sum_{n}\omega_{c,n}\hat{c}_{n}^{\dag}\hat{c}_{n} .
\end{equation}
Here, $\hat{a}_{l}^{\dag}$ and $\hat{a}_{l}$ are creation and annihilation operators for mode $l$ of reservoir $R_A$, with frequency $\omega_{a,l}$, and similarly for $R_B$ and $R_C$. The interaction between the qubits and reservoirs is given by
\begin{eqnarray}
\label{H-coupling}
\hat{H}_{SR} &=& \sum_{l}g_{A,l}\left(\hat{\sigma}_{+}^{A}\hat{a}_{l}+\hat{\sigma}_{-}^{A}\hat{a}_{l}^{\dag}\right)
+\sum_{m}g_{B,m}\left(\hat{\sigma}_{+}^{B}\hat{b}_{m}+\hat{\sigma}_{-}^{B}\hat{b}_{m}^{\dag}\right) \nonumber\\
&&+\sum_{n}\left[\left(g_{A,n}\hat{\sigma}_{+}^{A}+g_{B,n}\hat{\sigma}_{+}^{B}\right)\hat{c}_{n}+
\left(g_{A,n}\hat{\sigma}_{-}^{A}+g_{B,n}\hat{\sigma}_{-}^{B}\right)\hat{c}_{n}^{\dag}\right]
\end{eqnarray}
where $g_{A,l}$, $g_{B,m}$ are the coupling strengths of qubit $A$, $B$ with mode $l$, $m$ of reservoir $R_{A}$, $R_{B}$ respectively, while $g_{A,n}$ and $g_{B,n}$ denote that of qubit $A$ and $B$ respectively with mode $n$ of $R_{C}$. Here, we have used the rotating wave approximation in Eq. (\ref{H-coupling}) since the system-reservoir coupling strengthes are assumed to be much smaller than the system energy scale.

Based on the model given by $\hat{H}_S$, $\hat{H}_R$, and $\hat{H}_{SR}$, we proceed to construct a master equation for the evolution of the system qubits in the presence of the thermal reservoirs. We will work in the regime of weak system-baths interaction, where all system transition frequencies are large compared to the bath couplings. The reservoirs then couple to the delocalized eigenstates of the total system Hamiltonian $H_S$, and we will obtain a global master equation where each reservoir affects both qubits. For weak inter-system coupling one should instead employ a local master equation when each qubit is affected only by its local baths, as used e.g.~in Ref.~\cite{Brask2015}. The global approach is valid as long as the secular approximation holds, as detailed in \cite{Hofer2017} where the validity regime for local and global master equations for a thermal machine of two qubits or two harmonic oscillator was studied.

We construct the master equation in the basis of the eigenstates $\hat{H}_{S}$. In terms of the free Hamiltonian eigenstates, i.e.~$\ket{\eta_1} = \ket{11}$, $\ket{\eta_2} = \ket{10}$, $\ket{\eta_3} = \ket{01}$, and $\ket{\eta_4} = \ket{00}$, the eigenstates of $\hat{H}_{S}$ can be expressed as
$\ket{\lambda_1} = \ket{\eta_1}$,
$\ket{\lambda_2} = \cos\frac{\theta }{2} \ket{\eta_2} + \sin\frac{\theta }{2} \ket{\eta_3}$,
$\ket{\lambda_3} = - \sin\frac{\theta}{2} \ket{\eta_2} + \cos\frac{\theta }{2} \ket{\eta_3}$, and
$\ket{\lambda_4} = \ket{\eta_4}$,
and the corresponding eigenvalues as
$E_{1} = \epsilon_{A} + \epsilon_{B}$,
$E_{2} = \epsilon_{m} + \sqrt{\Delta\epsilon^2/4 + \Omega^2}$ ,
$E_{3} = \epsilon_{m} - \sqrt{\Delta\epsilon^2/4 + \Omega^2}$,
$E_{4} = 0$ with $\epsilon_{m}=(\epsilon_{A} + \epsilon_{B})/2$
and $\Delta\epsilon=\epsilon_{A} - \epsilon_{B} $.
The parameter $\theta$ is defined by $ \tan \theta = 2\Omega/ \Delta \epsilon $.

In terms of eigenstates of $\hat{H}_{S}$, the total Hamiltonian $\hat{H}_{tot}=\hat{H}_{S}+\hat{H}_{R}+\hat{H}_{SR}$ can be rewritten using
\begin{equation}
\label{HSdiag}
\hat{H}_{S}=\sum_{i=1}^{4}E_{i}\left|\lambda _{i}\right\rangle\left\langle\lambda _{i}\right| , \hspace{0.7cm} \mathrm{and} \hspace{0.7cm} \hat{H}_{SR} = \sum_{j=1}^{2}\hat{\widetilde{H}}_{SR,j} ,
\end{equation}
where
\begin{eqnarray}\label{HSRj}
\hat{\widetilde{H}}_{SR,j}&=&\sum_{l}g_{A,l}\left(\hat{V}^{+}_{A,j}\hat{a}_{l}+\hat{V}_{A,j}\hat{a}_{l}^{\dag}\right)
+\sum_{m}g_{B,m}\left(\hat{V}^{+}_{B,j}\hat{b}_{m}\right.\nonumber\\
&&\left.+\hat{V}_{B,j}\hat{b}_{m}^{\dag}\right)
+\sum_{n}\left[\left(g_{A,n}\hat{V}^{+}_{A,j}+g_{B,n}\hat{V}^{+}_{B,j}\right)\hat{c}_{n}\right. \\
&&\left.+\left(g_{A,n}\hat{V}_{A,j}+g_{B,n}\hat{V}_{B,j}\right)\hat{c}_{n}^{\dag}\right] . \nonumber
\end{eqnarray}
In this expression, $\hat{V}_{\mu,j}$ and $\hat{V}^{\dagger}_{\mu,j}$ are jump operators corresponding respectively to processes where the system looses an excitation to a bath or receives one from it. They are eigenoperators of $\hat{H}_{S}$, such that $[\hat{H}_{S},\hat{V}_{\mu,j}]=-\omega_{j}\hat{V}_{\mu,j}$ where the eigenfrequencies $\omega_{j}$ determine the energy lost or recieved by the system. They are given by $\omega_1 = E_3 - E_4 = E_1 - E_2 =\epsilon_{m} - \sqrt{\Delta\epsilon^2/4 + \Omega^2}$, corresponding to transitions $\ket{\lambda_1}\leftrightarrow\ket{\lambda_2}$ and $\ket{\lambda_3}\leftrightarrow\ket{\lambda_4}$, and $\omega_2 = E_1 - E_3 = E_2 - E_4 = \epsilon_{m} + \sqrt{\Delta\epsilon^2/4 + \Omega^2}$ corresponding to transitions $\ket{\lambda_1}\leftrightarrow\ket{\lambda_3}$ and $\ket{\lambda_2}\leftrightarrow\ket{\lambda_4}$. Explicitly, the $\hat{V}_{\mu,j}$ are constructed as follows
\begin{eqnarray}
\hat{V}_{A,1} &=& \sin \frac{\theta }{2}(\left| \lambda _{2} \right\rangle \left\langle {{\lambda _1}} \right| - \left| {{\lambda _4}} \right\rangle \left\langle {{\lambda _3}} \right|) , \nonumber \\
{\hat{V}_{A,2}} &=& \cos \frac{\theta }{2}(\left| {{\lambda _3}} \right\rangle \left\langle {{\lambda _1}} \right| + \left| {{\lambda _4}} \right\rangle \left\langle {{\lambda _2}} \right|) , \nonumber \\
{\hat{V}_{B,1}} &=& \cos \frac{\theta }{2}(\left| {{\lambda _2}} \right\rangle \left\langle {{\lambda _1}} \right| + \left| {{\lambda _4}} \right\rangle \left\langle {{\lambda _3}} \right|) , \nonumber \\
{\hat{V}_{B,2}} &=& \sin \frac{\theta }{2}( - \left| {{\lambda _3}} \right\rangle \left\langle {{\lambda _1}} \right| + \left| {{\lambda _4}} \right\rangle \left\langle {{\lambda _2}} \right|).
\end{eqnarray}

With the jump operators $\hat{V}_{\mu,j}$, one can derive a master equation on standard Lindblad form in the Born-Markov regime of weak coupling to the thermal reservoirs combined with a secular approximation, valid for strong inter-system coupling. Details of deriving a master equation from the system, bath, and interaction Hamiltonians can be found e.g.~in Ref.~\cite{def-curr1} Chap. 3 and Refs.~\cite{Hofer2017,Rivas2010,schaller2015}. A derivation with a common reservoir, as considered here, can be found in \cite{equi7}. One assumes a stationary state of the baths -- i.e. that the baths are sufficiently large to remain unaffected by the interaction with the system -- and neglects the Lamb shift, which is small compared to the qubit-qubit coupling strength $\Omega$ \cite{def-curr1,Rivas2010,NC} (see also the Appendix). We arrive at
\begin{equation}
\label{mast}
\dot\rho=-i[\hat{H}_{S},\rho]+\mathcal{L}_{A}[\rho]+\mathcal{L}_{B}[\rho]+\mathcal{L}_{C}[\rho],
\end{equation}
where $\mathcal{L}_{A}[\rho]$, $\mathcal{L}_{B}[\rho]$, and $\mathcal{L}_{C}[\rho]$ describe the dissipative effect on the qubits' dynamics due to coupling with the reservoirs $R_{A}$, $R_{B}$, and $R_{C}$ respectively. We note that, as we are working in the Born-Markov regime, the dissipators are additive \cite{Kolodynski2017} and so we can obtain the dynamics in the absence of $R_C$ simply by omitting the last term above. The dissipators arising from the independent baths are given by
\begin{eqnarray}
\label{LA}
\mathcal{L}_{A}[\rho] &=& \sum_{j} \Gamma_{A}(\omega_{j})\left[ \left(\overline{n}_{A}(\omega_{j}) + 1\right)\left(2\hat{V}_{A,j}\rho \hat{V}_{A,j}^{\dagger} -\left\{\hat{V}_{A,j}^{\dagger}\hat{V}_{A,j},\rho\right\}\right)\right. \nonumber \\
&& \left. + \overline{n}_{A}(\omega_{j})\left(2\hat{V}_{A,j}^{\dagger} \rho \hat{V}_{A,j} - \left\{\hat{V}_{A,j}\hat{V}_{A,j}^{\dagger},\rho\right\}\right)\right] ,
\end{eqnarray}
and
\begin{eqnarray}
\label{LB}
\mathcal{L}_{B}[\rho] &=& \sum_{j} \Gamma_{B}(\omega_{j})\left[ \left(\overline{n}_{B}(\omega_{j}) + 1\right)\left(2\hat{V}_{B,j}\rho \hat{V}_{B,j}^{+} -\left\{\hat{V}_{B,j}^{+}\hat{V}_{B,j},\rho\right\}\right) \right.\nonumber \\
&&\left. + \overline{n}_{B}(\omega_{j})\left(2\hat{V}_{B,j}^{+} \rho \hat{V}_{B,j} - \left\{\hat{V}_{B,j}\hat{V}_{B,j}^{+},\rho\right\}\right)\right].
\end{eqnarray}
In each case, the first line corresponds to stimulated and spontaneous emission, while the second line corresponds to absorption. $\Gamma_{A}(\omega_{j})$ and $\Gamma_{B}(\omega_{j})$, characterize the damping rates due to interactions with the reservoirs $R_{A}$ and $R_{B}$ respectively. Their exact forms depend on the spectral densities of the reservoirs. Each reservoir is assumed to be in a thermal state, and the occupation number (the average number of photons) at energy $\omega_j$ of reservoir $R_\nu$ ($\nu \in\{A,B,C\}$) is given by the Bose-Einstein distribution
\begin{equation}
\label{n}
\bar{n}_{\nu}(\omega_{j})=\frac{1}{\exp[\frac{\omega_{j}}{T_{\nu}}]-1}.
\end{equation}

In contrast to $R_{A}$ and $R_{B}$, the reservoir $R_{C}$ is common to the two qubits $A$ and $B$, and we see from (\ref{HSRj}) that it will introduce dissipative terms both of the forms (\ref{LA}) and (\ref{LB}) as well as cross terms. Therefore, we have $\mathcal{L}_{C}[\rho]=\mathcal{L}_{C}^{(A)}[\rho]+\mathcal{L}_{C}^{(B)}[\rho]
+\mathcal{L}_{C}^{(AB)}[\rho]$, in which $\mathcal{L}_{C}^{(A)}[\rho]$ and $\mathcal{L}_{C}^{(B)}[\rho]$ indicate dissipative effects due to qubit $A$ and $B$ coupling individually to $R_{C}$, while the term $\mathcal{L}_{C}^{(AB)}[\rho]$ reflects the collective coupling. Thanks to the collective effect of the common reservoir, the steady-state entanglement induced by independent reservoirs can be further enhanced. Explicitly
\begin{eqnarray}
\label{LCA}
\mathcal{L}_{C}^{(A)}[\rho] &=& \sum_{j} \Gamma_{C}^{(A)}(\omega_{j})\left[ \left(\overline{n}_{C}(\omega_{j}) + 1\right)\left(2\hat{V}_{A,j}\rho \hat{V}_{A,j}^{+} - \left\{\hat{V}_{A,j}^{+}\hat{V}_{A,j},\rho\right\}\right) \right.\nonumber \\
&&\left. + \overline{n}_{c}(\omega_{j})\left(2\hat{V}_{A,j}^{+} \rho \hat{V}_{A,j} - \left\{\hat{V}_{A,j}\hat{V}_{A,j}^{+},\rho\right\}\right)\right] , \\
\label{LCB}
\mathcal{L}_{C}^{(B)}[\rho] &=& \sum_{j} \Gamma_{C}^{(B)}(\omega_{j})\left[ \left(\overline{n}_{C}(\omega_{j}) + 1\right)\left(2\hat{V}_{B,j}\rho \hat{V}_{B,j}^{+} - \left\{\hat{V}_{B,j}^{+}\hat{V}_{B,j},\rho\right\}\right) \right.\nonumber \\
&&\left. + \overline{n}_{C}(\omega_{j})\left(2\hat{V}_{B,j}^{+} \rho \hat{V}_{B,j} - \left\{\hat{V}_{B,j}\hat{V}_{B,j}^{+},\rho\right\}\right)\right] ,
\end{eqnarray}
and
\begin{eqnarray}
\label{LCAB}
\mathcal{L}_{C}^{(AB)}[\rho] &=& \sum_{j} \Gamma_{C}^{(AB)}(\omega_{j})\left[ \left(\overline{n}_{C}(\omega_{j}) + 1\right)\left(2\hat{V}_{A,j}\rho \hat{V}_{B,j}^{+} - \left\{\hat{V}_{B,j}^{+}\hat{V}_{A,j},\rho\right\}\right) \right.\nonumber \\
&& + \overline{n}_{C}(\omega_{j})\left(2\hat{V}_{A,j}^{+}\rho \hat{V}_{B,j} - \left\{\hat{V}_{B,j}\hat{V}_{A,j}^{+},\rho\right\}\right) \nonumber \\
&& + \left(\overline{n}_{C}(\omega_{j}) + 1\right)\left(2\hat{V}_{B,j}\rho \hat{V}_{A,j}^{+} - \left\{\hat{V}_{A,j}^{+}\hat{V}_{B,j},\rho\right\}\right) \nonumber \\
&& \left.+ \overline{n}_{C}(\omega_{j})\left(2\hat{V}_{B,j}^{+}\rho \hat{V}_{A,j} - \left\{\hat{V}_{A,j}\hat{V}_{B,j}^{+},\rho\right\}\right)   \right] .
\end{eqnarray}
The collective damping rate fulfils $\Gamma_{C}^{(AB)}(\omega_{j})=\sqrt{\Gamma_{C}^{(A)}(\omega_{j})\Gamma_{C}^{(B)}(\omega_{j})}$.
For simplicity, in the remainder of the paper we will suppose that all the spectral densitites can be taken to be flat in the relevant energy range such that the damping rates are frequency independent, $\Gamma_{A}(\omega_{j})=\Gamma_{A}$, $\Gamma_{B}(\omega_{j})=\Gamma_{B}$, $\Gamma_{C}^{A}(\omega_{j})=\Gamma_{C}^{(A)}$ and $\Gamma_{C}^{(B)}(\omega_{j})=\Gamma_{C}^{(B)}$.

We are interested in the steady-state entanglement between the two qubits. The steady state is found by setting the left hand side of Eq.~(\ref{mast}) to zero, i.e.~by solving $\dot{\rho}^{S}=0$. The entanglement of the resulting two-qubit state can then be quantified by the concurrence \cite{con}. We obtain the steady state in the eigenbasis of $H_S$ with the density matrix elements $\lambda_{ii^{\prime}}^{S}=\left\langle\lambda_{i}\right|\rho^{S}\left|\lambda_{i^{\prime}}\right\rangle$. The state can then be reexpressed in the eigenbasis of the free Hamiltonian with density matrix elements $\eta_{ii^{\prime}}^{S}=
\left\langle\eta_{i}\right|\rho^{S}\left|\eta_{i^{\prime}}\right\rangle$ using
\begin{eqnarray}\label{eta}
\eta_{11}^{S}&=&\lambda_{11}^{S}, \nonumber\\
\eta_{22}^{S}&=& \cos^2\frac{\theta}{2} \lambda_{22}^{S}+ \sin^2\frac{\theta}{2}\lambda_{33}^{S} , \nonumber\\
\eta_{33}^{S}&=& \sin^2\frac{\theta}{2}\lambda_{22}^{S}  + \cos^2\frac{\theta}{2}\lambda_{33}^{S}, \nonumber\\
\eta_{44}^{S}&=&\lambda_{44}^{S}, \nonumber\\
\eta_{23}^{S}&=&\eta_{32}^{S} = \frac{1}{2}\sin\theta(\lambda_{22}^{S}-\lambda_{33}^{S}).
\end{eqnarray}
The only mechanism, which can generate coherence, is the inter-qubit interaction described by the Hamiltonian (\ref{Hint}). In fact, if there were no interaction between the qubits, there would be no process generating off-diagonal terms in the free Hamiltonian eigenbasis. The bath-induced dissipation tends to destroy coherence. Thus, only coherences induced by the interaction can survive in the steady state, and $\rho^S$ will therefore be of the form
\begin{equation}\label{X}
\rho^{S}=\left(
       \begin{array}{cccc}
         \eta_{11}^{S} & 0 & 0 & 0 \\
         0 & \eta_{22}^{S} & \eta_{23}^{S} & 0 \\
         0 & \eta_{32}^{S} & \eta_{33}^{S} & 0 \\
         0 & 0 & 0 & \eta_{44}^{S} \\
       \end{array}
     \right) .
\end{equation}
This is a so-called `X state' for which the concurrence reduces to the simple expression \cite{xtype}
\begin{equation}
\mathcal{C}(\rho^{S})=2 \max\{0,|\eta_{23}^{S}|-\sqrt{\eta_{11}^{S}\eta_{44}^{S}}\}.
\end{equation}
The state is entangled whenever $\mathcal{C}(\rho^{S}) > 0$ and maximally entangled for  $\mathcal{C}(\rho^{S}) = 1$

In addition to the entanglement, it also interesting to look at the heat currents in the system. The introduction of a common reservoir with its own associated temperature will influence both the entanglement and heat current, and we will investigate this link below. The heat current associated with reservoir $R_{\nu}$ can be defined as\footnote{Note that while a master equation of global type and employing a secular approximation, as we do here, can lead to zero predictions for the inter-system energy current between the qubits \cite{Wichterich2007}, there are no inconsistencies associated with evaluating the currents between the system and reservoirs \cite{Hofer2017}.} \cite{def-curr1,def-curr2}
\begin{equation}  \label{current}
Q_{\nu}=Tr\{\mathcal{L}_{\nu}[\rho^{S}]\hat{H}_{S}\}.
\end{equation}
In Eq. (\ref{current}), $\mathcal{L}_\nu[\rho^S]$ represents the change in the system state induced by the bath $\nu$ and the trace of it with the system Hamiltonian hence represents the associated change in the system energy.
From the perspective of reservoir, a positive heat current means heat release from the reservoir, while a negative value implies heat absorption by the reservoir. Therefore, a sign change of the heat current indicates a crossover between heat absorption and heat release or vice versa.

\section{Results}

We now analyse how steady-state entanglement generation and heat currents are influenced by the introduction of the common heat reservoir $R_C$. We first consider the case where the two independent reservoirs $R_A$ and $R_B$ are in thermal equilibrium, i.e.~$T_A=T_B$, and then turn to the out-of-equilbrium case below. We will compare the amount of steady-state entanglement with $R_C$ present with the amount when the system is decoupled from $R_C$, and also examine the heat currents.

\subsection{Independent reservoirs at thermal equilibrium}

\begin{figure}[tbp]
\begin{center}
{\includegraphics[width=0.7\linewidth]{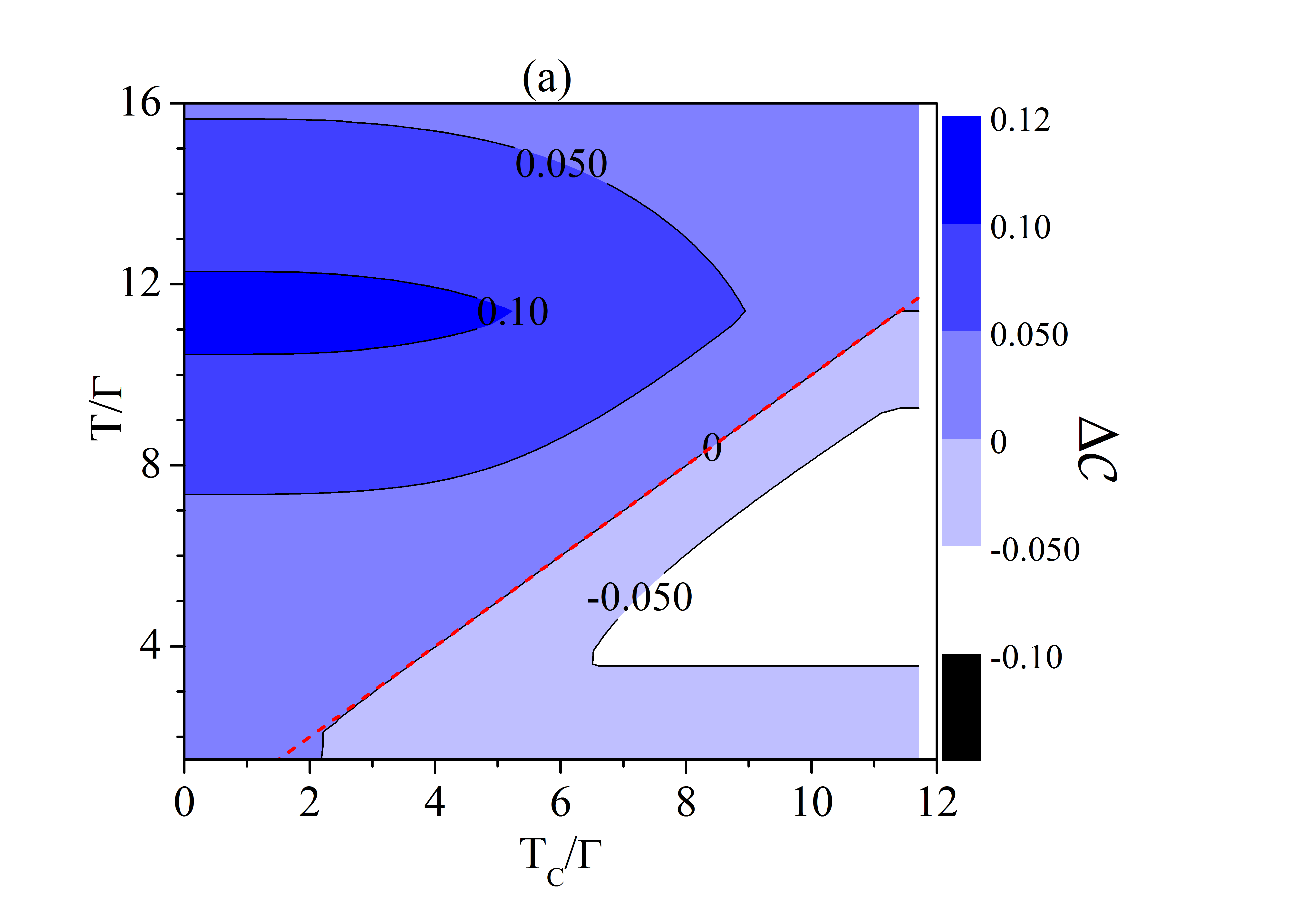} %
\includegraphics[width=0.7\linewidth]{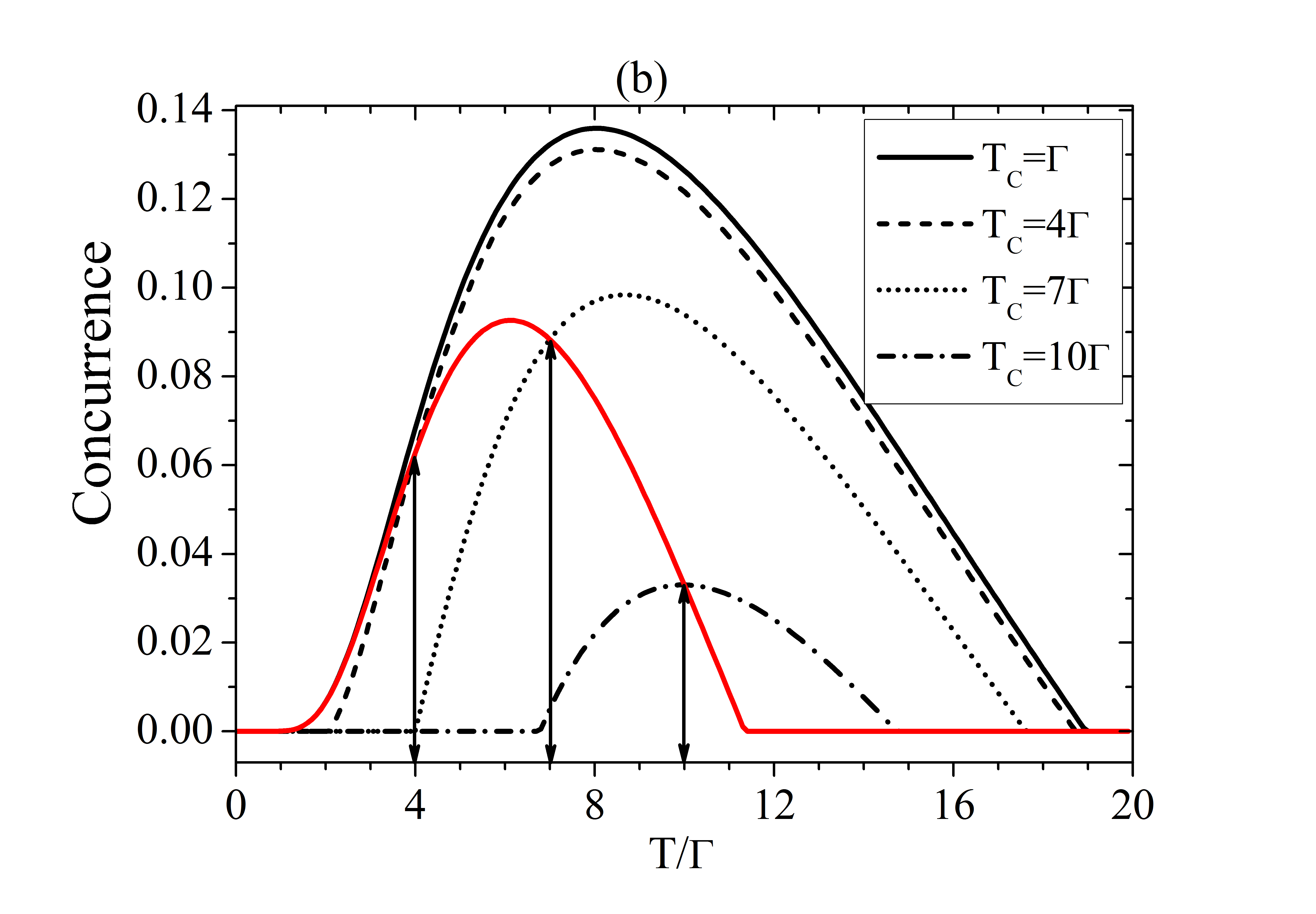} %
}
\end{center}
\caption{(Color online) \textbf{(a)} Difference in steady-state concurrence $\Delta\mathcal{C}$ with and without the common bath vs.~the temperatures $T=T_A=T_B$ and $T_{C}$. The red dashed curve indicates $T_{C}=T$ where $\Delta\mathcal{C}=0$. \textbf{(b)} Steady-state concurrence vs.~$T$ without the common reservoir $\mathcal{C}_{AB}$ (red curve) and with $\mathcal{C}_{ABC}$ for different $T_{c}$ (black curves). Arrows label the crossing points where $T=T_C$. In both plots, the remaining parameters are given by $\Gamma_{A}=\Gamma_{B}=\Gamma_{C}^{(A)}=\Gamma_{C}^{(B)}=\Gamma$, $\epsilon_{A}=\epsilon_{B}=20\Gamma$, and $\Omega=10\Gamma$. }
\label{ent-equi}
\end{figure}

In this section, we consider $R_{A}$ and $R_{B}$ to be in the thermal equilibrium with $T_{A}=T_{B}=T$. We will also focus on the case where the qubits are resonant, $\epsilon_A=\epsilon_B=\epsilon$ (i.e.~$\theta=\pi/2$). In the absence of the third reservoir $R_C$, the system will relax into a thermal equilibrium state with temperature $T$ which may contain thermal entanglement \cite{equi1,equi2,equi3,equi4,equi5,equi6,equi7}. We are interested in how the amount of entanglement varies when the common reservoir with temperature $T_{C}$ is introduced and the entanglement depends on $T$ and $T_{C}$. That is, denoting the concurrence what the system is decoupled from $R_C$ by $\mathcal{C}_{AB}$ and the concurrence in the presence of $R_C$ by $\mathcal{C}_{ABC}$, we want to compare $\mathcal{C}_{AB}(T)$ with $\mathcal{C}_{ABC}(T,T_C)$.

\begin{figure}[tbp]
\begin{center}
{\includegraphics[width=0.7\linewidth]{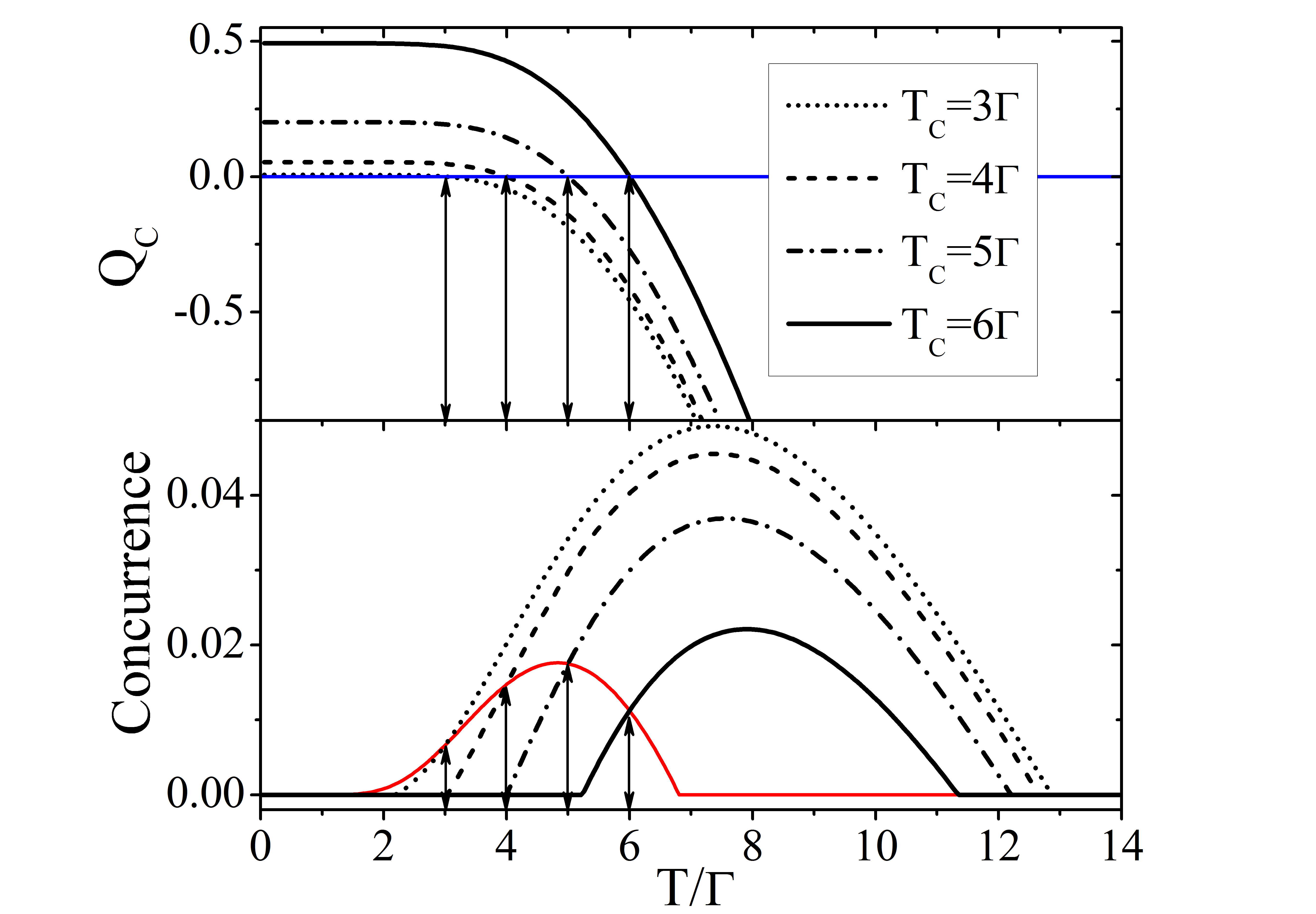} }
\end{center}
\caption{(Color online) Heat current $Q_{C}$ out of the common reservoir $R_{C}$ (top panel) and steady-state concurrence (bottom panel) vs.~$T_{A}=T_{B}=T$
for different $T_{c}$ (black curves). The red line indicates the concurrence in
the absence of $R_C$. Arrows label the points where $T=T_C$. The remaining parameters are $\Omega=6\Gamma$, $\epsilon_{A}=\epsilon_{B}=20\Gamma$, $\Gamma_{A}=\Gamma_{B}=\Gamma_{C}^{(A)}=\Gamma_{C}^{(B)}=\Gamma$.}
\label{ent-heat-equi}
\end{figure}

In Fig.\ref{ent-equi} (a), we plot the difference $\Delta\mathcal{C} = \mathcal{C}_{ABC}(T,T_C) - \mathcal{C}_{AB}(T)$ as a function of the temperatures. This is the change in steady-state entanglement induced by introducing the common reservoir at temperature $T_C$. As might be expected, we observe that when $T_{C}=T$ (red, dashed line in the figure) there is no change, because the system retains the same thermal equilibrium state with temperature $T$. The concurrence increases when $T_C < T$, while it decreases for $T_C > T$. Thus, the common reservoir enhances the steady-state entanglement when it effectively cooling the system. In Fig.\ref{ent-equi} (b), we plot the steady-state entanglement as a function of $T$ without the common reservoir, i.e.~$\mathcal{C}_{AB}(T)$ as well as with, $\mathcal{C}_{ABC}(T,T_C)$ for different $T_C$. We clearly see that the maximum of $\mathcal{C}_{ABC}$ can be significantly higher than that of $\mathcal{C}_{AB}$. While entanglement vanishes for large $T$ without $R_C$, it can be recovered by adding the common reservoir. When the common reservoir is cold (low $T_C$), the steady-state entanglement can retain a finite nonzero value for larger $T$, indicating that the thermal gradient induced by different $T$ and $T_C$ assists the entanglement generation. This can be further corroborated by studying the heat current $Q_{C}$ out of the reservoir $R_{C}$, which we plot in Fig.~\ref{ent-heat-equi}. When $T_C$ goes from being larger than $T$ to being smaller, the current changes sign from positive to negative meaning that the reservoir $R_{C}$ begins to absorb heat. The enhancement of steady-state entanglement is thus accompanied by heat absorption of the common reservoir from the independent ones.

\begin{figure}[tbp]
\begin{center}
{\includegraphics[width=0.7\linewidth]{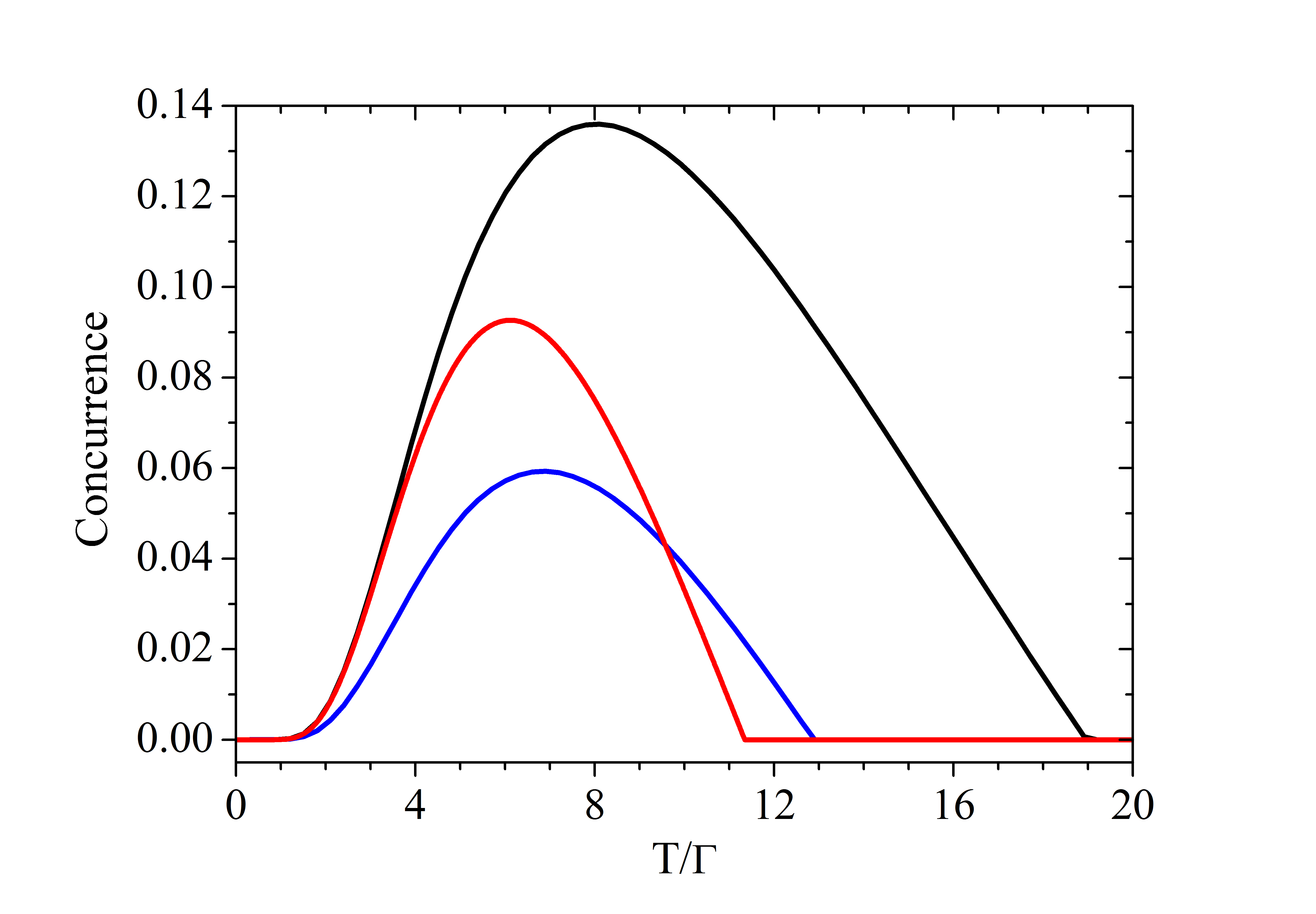} }
\end{center}
\caption{(Color online) A comparison of the steady-state concurrence vs.~$T=T_{A}=T_{B}$ when the collective dissipator $\mathcal{L}_{C}^{(AB)}[\rho]$ (\ref{LCAB}) is removed from the qubits' dynamics (blue curve) to that with it (black curve) as well as to that when the common reservoir is decoupled (red curve).
The parameters are set as $T_{C}=\Gamma$, $\epsilon_{A}=\epsilon_{B}=20\Gamma$, $\Omega=10\Gamma$
and $\Gamma_{A}=\Gamma_{B}=\Gamma_{C}^{(A)}=\Gamma_{C}^{(B)}=\Gamma$.}
\label{comp}
\end{figure}

The enhancement of entanglement is due to the collective effect of the common reservoir, represented by the collective dissipator $\mathcal{L}_{C}^{(AB)}[\rho]$ in Eq. (\ref{LCAB}). To visualise this, in Fig.~\ref{comp} we compare the steady-state concurrence when $\mathcal{L}_{C}^{(AB)}[\rho]$ is removed from the qubits' dynamics (blue curve) to that with it (black curve) as well as to that when the common reservoir is completely decoupled (red curve). Clearly, in absence of $\mathcal{L}_{C}^{(AB)}[\rho]$, the introduction of a common reservoir instead suppresses the concurrence for most $T$ compared to the situation without a common reservoir. Hence we see that it is the collective action which generates the entanglement enhancement. As we have shown above, to efficiently exert the collective effect, the temperature of the common reservoir should be lower than that of the independent reservoirs in the thermal equilibrium case. Intuitively, the dissipators $\mathcal{L}_{C}^{(A)}$ and $\mathcal{L}_{C}^{(B)}$ have a similar effect as the independent reservoirs. If the common reservoirs is warmer than the independent ones, they tend to heat the system, making it more mixed and destroying entanglement. This effect competes with the enhancement induced by $\mathcal{L}_{C}^{(AB)}$. Also,
note that the enhancement cannot be obtained without the common bath by simple cooling using the individual baths, i.e. by lowering their temperature. This can be seen since it is also present for temperatures below the peak of the red curve in Fig. ~\ref{comp}.

\subsection{Independent reservoirs out of thermal equilibrium}

We now turn to the case where the two independent reservoirs $R_A$, $R_B$ are not necessarily at thermal equilibrium, $T_A \neq T_B$. In the regime of weak qubit-qubit interaction, where there is negligible entanglement at equilibrium, such a temperature gradient can be harnessed for entanglement generation, as shown for thermal machines \cite{Brunner2014,Brask2015,BraskPRE2015,Tavakoli2017}. In Refs.~\cite{Brask2015,Tavakoli2017} entanglement was maximised when the temperature difference was as large as possible, e.g.~for $T_A$ approaching zero and $T_B$ large. Here, we are interested in how the addition of a common reservoir $R_C$ affects the amount of steady-state entanglement. In particular, we saw above that in equilibrium $T_A=T_B=T$, the addition of $R_C$ enhances the entanglement whenever $T_C < T$. We would like to understand how this finding generalises to the nonequilibrium setting.

Since the qubit-qubit interaction is strong, where the reservoirs coupled to the delocalised eigenstates of the system Hamiltonian $H_S$, we can be regard our model as describing an effective four-level system connected with two independent reservoirs (in the absence of $R_C$). Out of equilibrium, the steady state of this system is not generally a Gibbs state, and so it is not possible to assign it a temperature in an unambiguous manner. Nevertheless, we can characterize the state of the effective four-level system via the following two effective temperatures \cite{equi7,Teff1,Teff2,Teff3} as
\begin{equation}
\label{Teff}
T_{eff}(\omega_{1})=\frac{\omega_{1}}{\ln(\Gamma_{1}^{-}/\Gamma_{1}^{+})}, \hskip 0.5cm
T_{eff}(\omega_{2})=\frac{\omega_{2}}{\ln(\Gamma_{2}^{-}/\Gamma_{2}^{+})},
\end{equation}
where
\begin{eqnarray}
\label{Gammaspm}
\Gamma_{1}^{-} &=& \sin^{2}(\frac{\theta}{2})\Gamma_{A}(\omega_{1})[\overline{n}_{A}(\omega_{1})+1]
+\cos^{2}(\frac{\theta}{2})\Gamma_{B}(\omega_{1})[\overline{n}_{B}(\omega_{1})+1], \nonumber \\
\Gamma_{1}^{+} &=& \sin^{2}(\frac{\theta}{2})\Gamma_{A}(\omega_{1})\overline{n}_{A}(\omega_{1})
+\cos^{2}(\frac{\theta}{2})\Gamma_{B}(\omega_{1})\overline{n}_{B}(\omega_{1}), \nonumber\\
\Gamma_{2}^{-} &=& \cos^{2}(\frac{\theta}{2})\Gamma_{A}(\omega_{2})[\overline{n}_{A}(\omega_{2})+1]
+\sin^{2}(\frac{\theta}{2})\Gamma_{B}(\omega_{2})[\overline{n}_{B}(\omega_{2})+1], \nonumber\\
\Gamma_{2}^{+} &=& \cos^{2}(\frac{\theta}{2})\Gamma_{A}(\omega_{2})\overline{n}_{A}(\omega_{2})
+\sin^{2}(\frac{\theta}{2})\Gamma_{B}(\omega_{2})\overline{n}_{B}(\omega_{2}),
\end{eqnarray}
denote effective transition rates between the eigenstates of $H_S$ (see e.g.~\cite{equi7} for details). When the two independent reservoirs are in thermal equilibrium, $T_{A}=T_{B}=T$, both effective temperatures reduce to $T$, consistent with the fact that the two coupled qubits eventually reach a thermal equilibrium state. By contrast, out of equilibrium, $T_{A}\neq T_{B}$, the effective temperatures are generally different, both in the range between $\min\{T_A,T_B\}$ and $\max\{T_A,T_B\}$. Depending on the reservoir temperature and the detuning between the qubits $\Delta\epsilon$, one or the other effective temperature may be larger. There thus exist some special conditions under which the two effective temperatures become equal even when reservoirs $R_A$ and $R_B$ are not in equilibrium, in which case we \textit{can} assign a definite temperature to the system. We will now see that in these special situations, the result obtained in the equilibrium case above still holds.

\begin{figure}[tbp]
\begin{center}
{\includegraphics[width=0.7\linewidth]{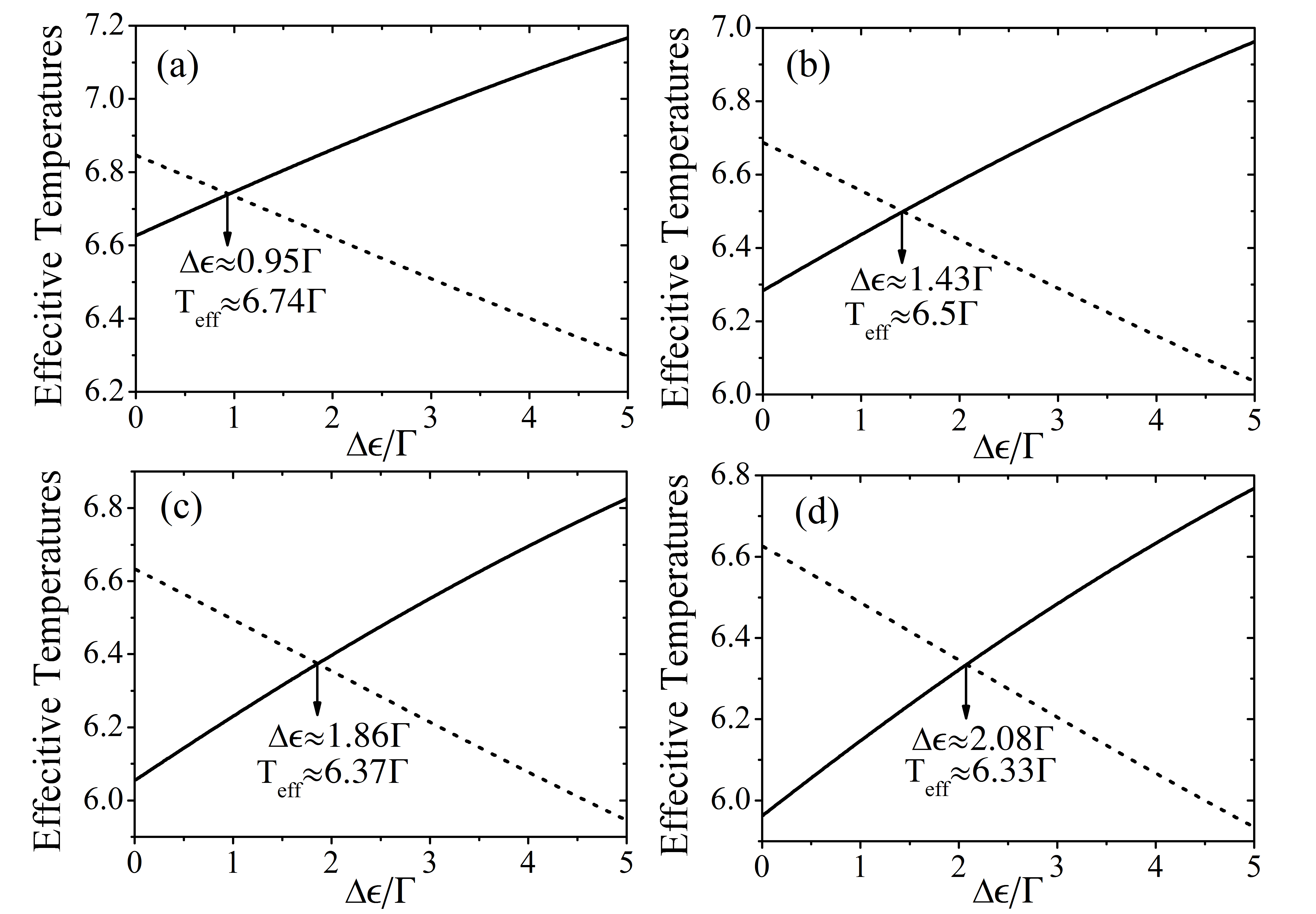} }
\end{center}
\caption{Plot of the effective temperatures $T_{eff} (\omega_1)$ (solid lines) and $T_{eff} (\omega_2)$ (dashed lines) in the absence of the common reservoir vs.~the detuning $\Delta\epsilon/\Gamma$ for $T_{B}=8\Gamma$  and \textbf{(a)} $T_{A}=5\Gamma$, \textbf{(b)} $T_{A}=4\Gamma$, \textbf{(c)} $T_{A}=3\Gamma$, and \textbf{(d)} $T_{A}=2\Gamma$. The remaining parameters are $\Omega=6\Gamma$, $\epsilon_{m}=20\Gamma$, and $\Gamma_{A}=\Gamma_{B}=\Gamma$.}
\label{eff-tem}
\end{figure}

\begin{figure}[tbp]
\begin{center}
{\includegraphics[width=0.7\linewidth]{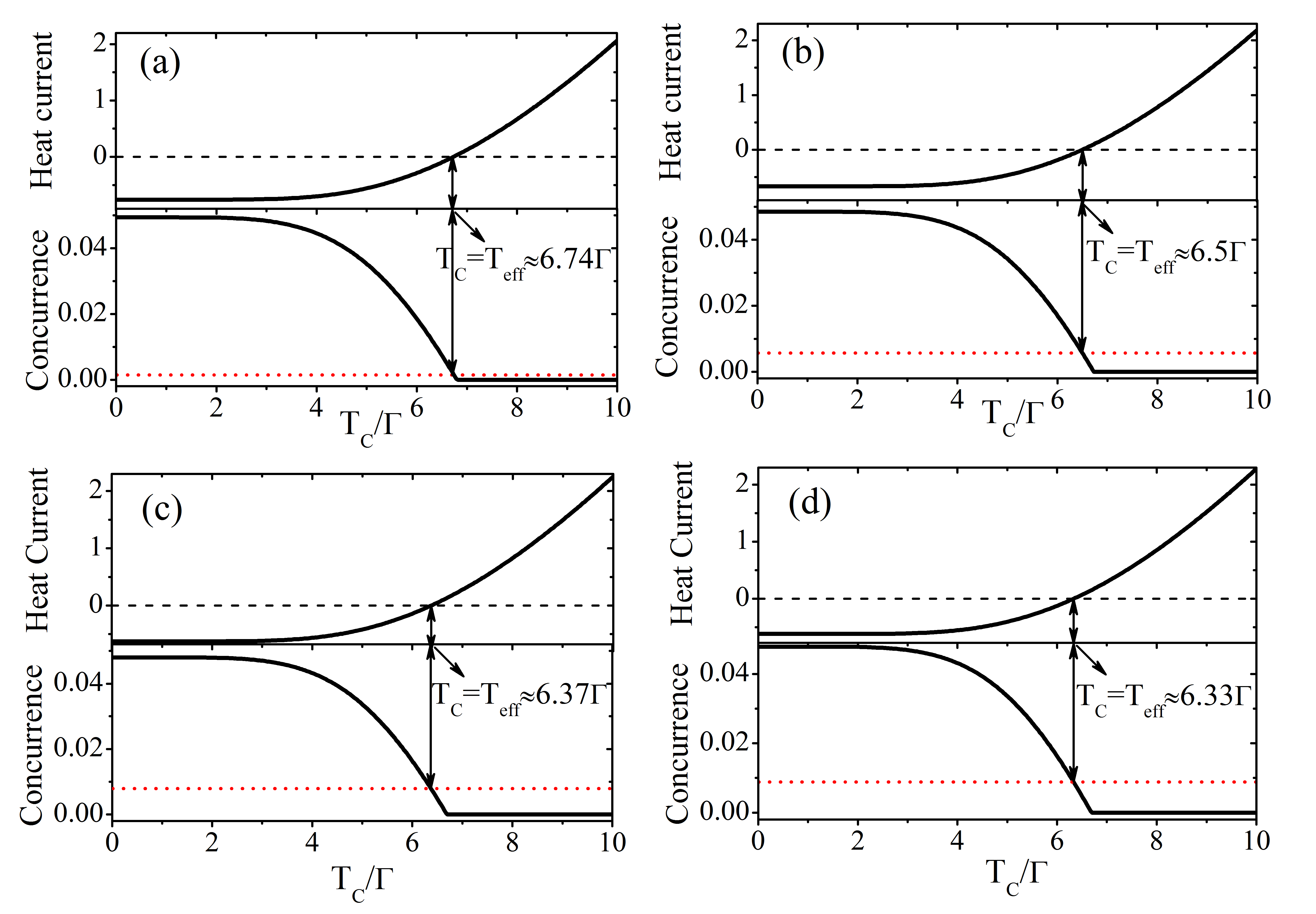}
}
\end{center}
\caption{(Color online) Each panel shows the heat current $Q_C$ (top) and concurrence (bottom) vs.~the temperature of the common reservoir $R_C$ for the thermalisation points found in Fig.~\ref{eff-tem}, namely \textbf{(a)} $T_{A}=5\Gamma$, $\Delta\epsilon=0.95\Gamma$, \textbf{(b)} $T_{A}=4\Gamma$, $\Delta\epsilon=1.43\Gamma$, \textbf{(c)} $T_{A}=3\Gamma$, $\Delta\epsilon=1.86\Gamma$, and \textbf{(d)} $T_{A}=2\Gamma$, $\Delta\epsilon=2.08\Gamma$. Red, dotted lines indicate the concurrence in the absence of $R_C$. Dashed lines indicate the zero point for the heat currents, and arrows mark the points where the current changes direction. At these points, the concurrence in the presence of $R_C$ is the same as that without it. When $T_{C}<T_{eff}$ the concurrence is enhanced due to the involvement of the common reservoir. The remaining parameters are the same as that in Fig. ~\ref{eff-tem}. }
\label{ent-heat-nonequi}
\end{figure}

From (\ref{Teff}) and (\ref{Gammaspm}), for a given temperature gradient, we can derive a suitable energy detuning $\Delta\epsilon$ of the two qubits such that $T_{eff}(\omega_{1})=T_{eff}(\omega_{2})=T_{eff}$. This is illustrated in Fig.~\ref{eff-tem}, where we plot the two effective temperatures $T_{eff}(\omega_{1})$ and $T_{eff}(\omega_{2})$ as functions of the detuning $\Delta\epsilon$ for different temperature gradients. For the points of thermalisation in Fig.~\ref{eff-tem}, where $T_{eff}(\omega_{1})=T_{eff}(\omega_{2})=T_{eff}$, we now consider the effect of adding the common reservoir $R_C$. In Fig.~\ref{ent-heat-nonequi} we show the concurrence and heat current $Q_{C}$ as functions of $T_{C}$. The values in the absence of $R_C$ are also indicated. We see that the amount of entanglement is enhanced with respect to that obtained in the absence of $R_C$ whenever $T_{C}<T_{eff}$. Thus our statement from the equilibrium case above generalises with $T$ replaced by $T_{eff}$. As before, the lower $T_C$ the higher the concurrence. Again, enhancement of entanglement is associated with heat absorption by the common reservoir ($Q_{C}$ becomes negative).

\begin{figure}[tbp]
\begin{center}
{\includegraphics[width=\linewidth]{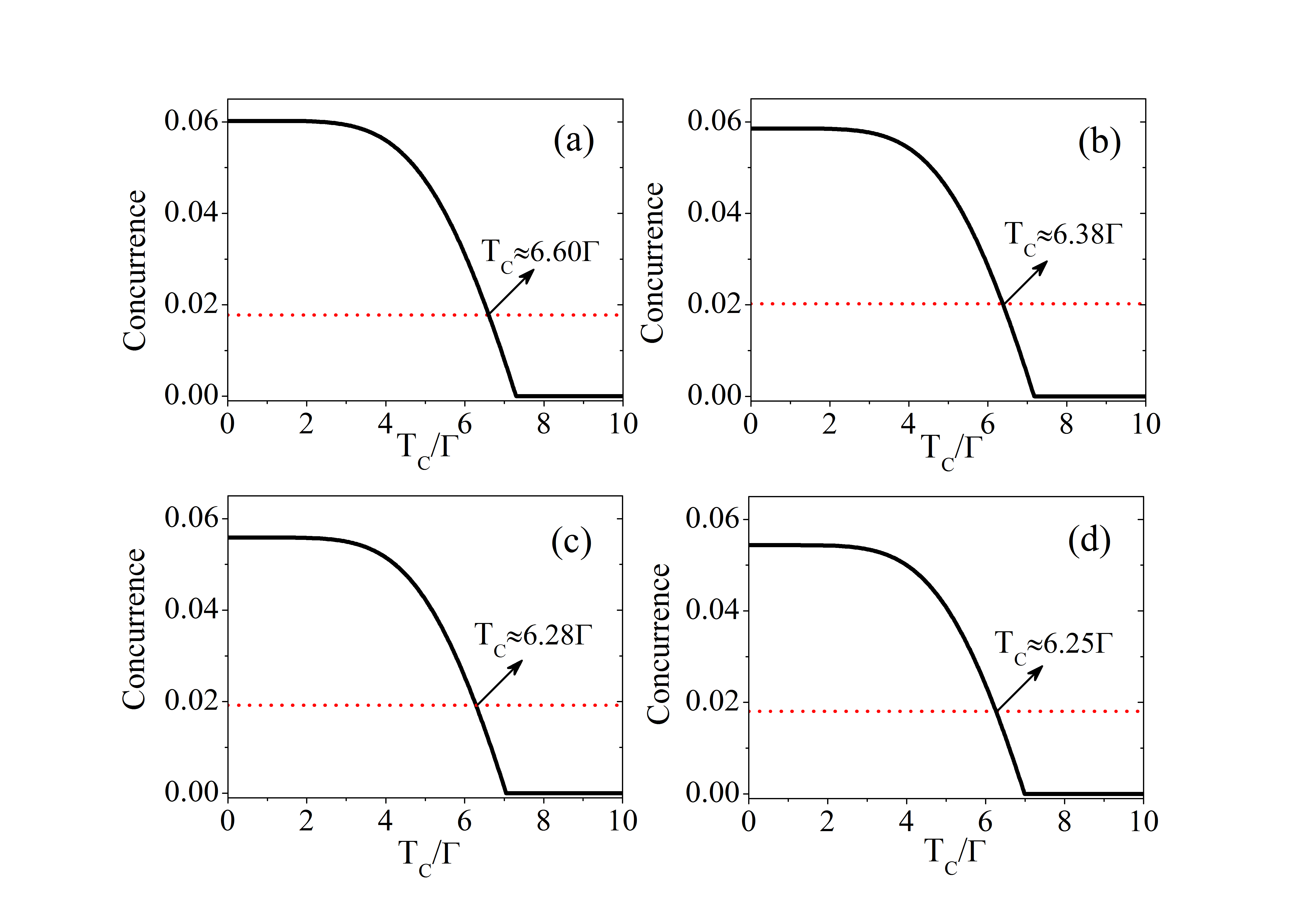}
}
\end{center}
\caption{(Color online) Each panel shows the concurrence vs.~the temperature of the common reservoir $R_C$ when an effective temperature cannot be assigned to the system. The parameters in the present four panels are the same as that in Figs.~\ref{eff-tem} and ~\ref{ent-heat-nonequi} but with a choice of $\Delta\epsilon=3\Gamma$ so that an effective temperature cannot be reached. The entanglement promotion can still be achieved when $T_{C}<6.60\Gamma\in\left\{T_{eff} (\omega_1)=6.97\Gamma,T_{eff} (\omega_2)=6.51\Gamma\right\}$ in (a), $T_{C}<6.38\Gamma\in\left\{T_{eff} (\omega_1)=6.72\Gamma,T_{eff} (\omega_2)=6.29\Gamma\right\}$ in (b), $T_{C}<6.28\Gamma\in\left\{T_{eff} (\omega_1)=6.55\Gamma,T_{eff} (\omega_2)=6.21\Gamma\right\}$ in (c) and $T_{C}<6.25\Gamma\in\left\{T_{eff} (\omega_1)=6.48\Gamma,T_{eff} (\omega_2)=6.20\Gamma\right\}$, namely, the temperature $T_{C}$ of the common reservoir is less than a value bing in between the two effective temperatures $T_{eff} (\omega_1)$ and $T_{eff} (\omega_2)$. }
\label{noeffT}
\end{figure}

For the particular nonequilibrium conditions under which the qubits can be assigned a single effective temperature, we have thus shown that entanglement can be improved for $T_{C}$ up to $T_{eff}$. In the general nonequilibrium case where there is no single thermalisation temperature, $T_{eff}(\omega_{1})\neq T_{eff}(\omega_{2})$, the addition of a common reservoir with suitable temperature can still improve the steady-state entanglement. Although we could not explicitly determine an upper bound on $T_C$ below which entanglement is increased in this general case, we numerically verify that such upper bound lies in between the two effective temperatures $T_{eff}(\omega_{1})$ and $T_{eff}(\omega_{2})$ of the system. To be visualized, in Fig.~\ref{noeffT} we exhibit the concurrence vs.~the temperature of the common reservoir $R_C$ for this case. Based on the findings in Fig.~\ref{eff-tem} on the relations of effective temperatures and the detuning $\Delta\epsilon$, here we choose $\Delta\epsilon=3\Gamma$ so that an effective temperature cannot be assigned to the system being contrary to the choice in Fig.~\ref{ent-heat-nonequi}. We can see from Fig.~\ref{noeffT} that the entanglement promotion can still be achieved when the temperature $T_{C}$ of the common reservoir is less than a value bing in between the two effective temperatures $T_{eff} (\omega_1)$ and $T_{eff} (\omega_2)$.

\section{Implementation}

Before we conclude, in this section we propose a possible implementation of our scheme in circuit quantum electrodynamics (QED) and compute the achievable improvement in steady-state entanglement for experimentally accessible values of the coupling parameters. A number of physical platforms could potentially enable implementations of the scheme, including trapped atoms, ions, and solid-state artificial atoms such as nitrogen-vacancy centres in diamond. However, here we focus on superconducting systems in which experimental studies of quantum thermodynamics have already been realised \cite{Cottet2017,Koski2013,Koski2014,Pekola2015} and which are good candidates for implementing quantum thermal machines \cite{Brask2015,Chen2012,Hofer2016a,Hofer2016b,Tavakoli2017}.

\begin{figure}[tbp]
\centering
\includegraphics[width=0.7\linewidth]{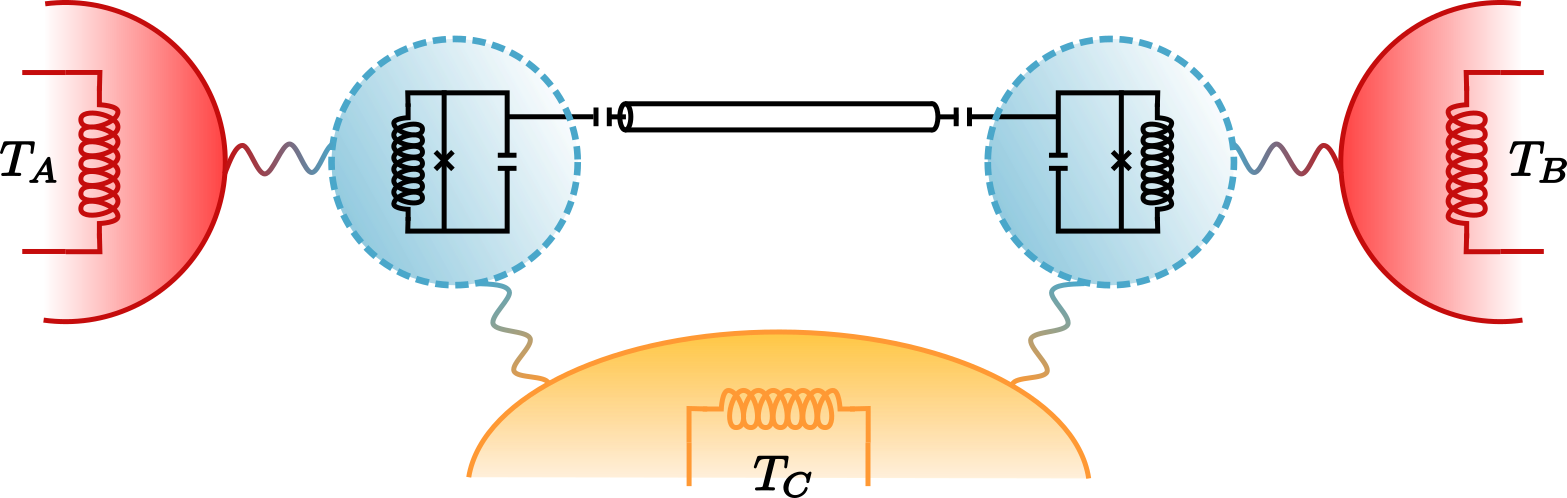}
\caption{Possible circuit QED implementation of the scheme. Two fluxonium qubits are coupled via a microwave resonater detuned from the energy spacing of the qubits, to realise a system Hamiltonian $H_{S}$ given in (\ref{H0})-(\ref{Hint}). Each qubit is coupled to effective baths with variable temperatures, corresponding to noise in external circuits which have a finite impedance. A common bath affecting both qubits is realised in the same manner. Imperfect control over external control parameters and other noise sources leads to additional pure dephasing.}
\label{fig.implementation}
\end{figure}

In circuit QED, a Hamiltonian of the form $\hat{H}_{S}$ can be realised by two transmon or fluxonium qubits \cite{Manucharyan2012}, as shown in Fig.~\ref{fig.implementation}. The level spacing of fluxonium qubits is accurately tunable in a wide range from hundreds of MHz to tens of GHz. Several coupling mechanisms are available for realising the qubit-qubit interaction. Both transmon qubits \cite{Majer2007,Sillanpaa2007,DiCarlo2009} and fluxonium qubits can be coupled capacitively or inductively via a cavity in the dispersive regime (of strong detuning of the qubits and cavity from the strength of the qubit-cavity coupling) \cite{Manucharyan2012}. Second, an alternative is direct mutual inductive coupling as described in \cite{Chen2014} and proposed for fluxonium qutrits in \cite{Manucharyan2012}. We note that while achieving strong inter-qubit coupling as considered here is certainly challenging, the dispersive regime is not a strict limitation  \cite{Filipp2011,Kim2015}. It only requires that the detuning between the qubit frequency and the cavity frequency be larger than each qubit-resonator frequency.

\begin{figure*}[tbp]
\includegraphics[width=\linewidth]{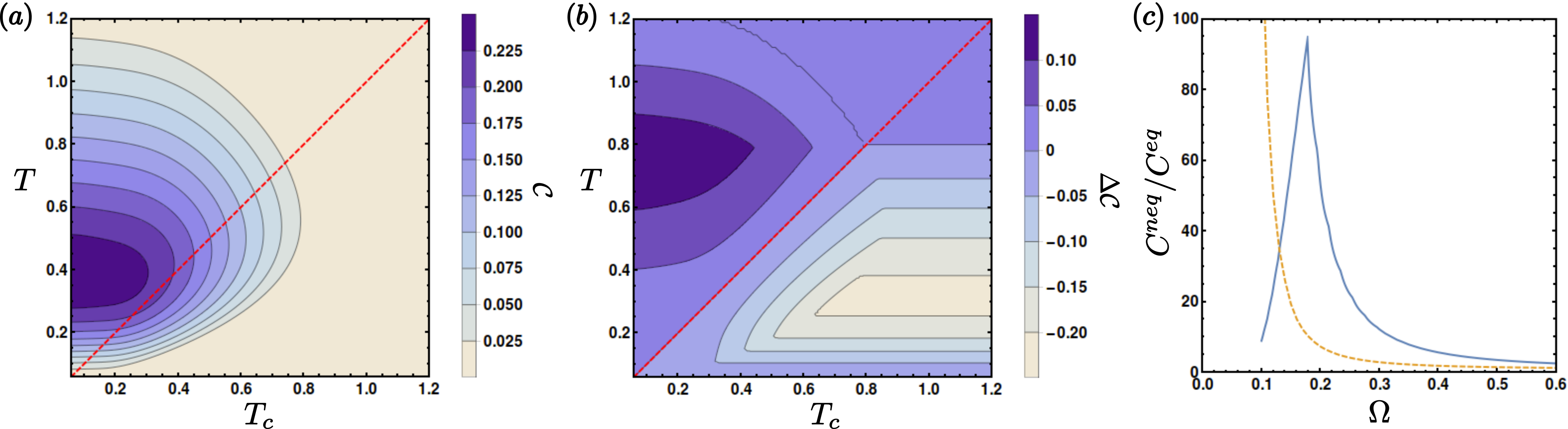}
\caption{\textbf{(a)} Steady-state entanglement as measured by the concurrence vs.~the temperatures of the individual (taken equal $T=T_A=T_B$) and common baths. The qubit transition and interaction frequencies are $\epsilon_A=\epsilon_B=1\,\mathrm{GHz}$ and $\Omega=0.7\,\mathrm{GHz}$. The bath coupling strengths are $\Gamma_A=\Gamma_B=\Gamma_C^{(A)}=\Gamma_C^{(B)}=10\,\mathrm{MHz}$, and the pure dephasing rate is $\gamma = 3.5\times 10^{-2}\,\mathrm{MHz}$. The red, dashed line indicates thermal equilibrium $T=T_C$. \textbf{(b)} The relative improvement in concurrence $\Delta\mathcal{C}$ for the same parameters as in (a). \textbf{(c)} Ratio of the maximal steady-state concurrence out of equilibrium $C^{neq}$ to the maximum in equilibrium $C^{eq}$ for the same energy gaps and pure dephasing rate as in (a). For the orange, dashed curve, the bath couplings are as in (a) while for the solid, blue curve $\Gamma_A=\Gamma_B=\Gamma_C^{(A)}=\Gamma_C^{(B)}=0.1\,\mathrm{MHz}$.}
\label{fig.implplots}
\end{figure*}

The qubits are naturally coupled to thermal baths due to the presence of thermal Johnson Nyquist noise in the surrounding circuitry. Effective thermal baths for each qubit can be implemented by controling the electronic noise coupling to each qubit. E.g.~the effective temperature can be increased by increasing the noise level in particular transmission lines. A common bath coupling to both qubits can be realised similarly. Specifically, a transmission line coupling to the cavity which mediates the qubit-qubit interaction will couple to both qubits and can provide a common bath. If we model the thermal environments of the qubits by bosonic thermal reservoirs, their effect on the system is already captured by the Lindblad-type master equation (\ref{mast}). The system-bath coupling strengths can vary in a range of about 0.1-10 MHz. Imperfections in external control parameters, such as magnetic flux noise, will lead to additional pure dephasing \cite{OMalley2016}. We account for this phenomenologically by adding another dissipative term on the right-hand-side of (\ref{mast}), given by
\begin{equation}
\mathcal{L}_{dep}[\rho] = \gamma( \hat{D}_A \rho \hat{D}_A^\dagger + \hat{D}_B \rho \hat{D}_B^\dagger - 2\rho)
\end{equation}
where $\hat{D}_A = \hat{\sigma}_z\otimes\mathbb{I}$ and $\hat{D}_B = \mathbb{I}\otimes\hat{\sigma}_z$ with $\hat{\sigma}_z$ the Pauli operator, and $\gamma$ is the pure dephasing rate which we take to be the same for both qubits. Based on the relaxation ($T_1$) and Ramsey dephasing ($T_2$) times measured for fluxonium qubits in Ref.~\cite{Kou2017}, we take the pure dephasing rate to be $\gamma = T_2^{-1} - (2T_1)^{-1} \approx 3.5\times 10^{-2} \,\mathrm{MHz}$.

Solving for the steady state of the modified master equation, we can compute the attainable concurrence for experimentally relevant parameter settings. We obtain the result shown in Fig.~\ref{fig.implplots}, where the temperatures are given in units of the qubit transition frequency, which is set to 1 GHz. As can be seen from Fig.~\ref{fig.implplots}(a), a significant amount of entanglement can be generated in a realistic setting, even in the presence of pure dephasing. From Fig.~\ref{fig.implplots}(b) we also note that the conclusion from above is still valid: More entanglement can be generated when the common bath temperature is below the individual bath temperatures, and hence the system is out of equilibrium. The improvement in steady-state entanglement depends on the qubit-qubit interaction strength, as well as the strength of the bath couplings, as shown in Fig.~\ref{fig.implplots}(c). When the bath coupling is weaker, the improvement peaks at higher interaction strengths. Substantial improvements can be obtained for accessible parameter values.

\section{Conclusion}

In conclusion, we have shown that it is possible to improve the steady-state entanglement of two interacting qubits coupled to independent thermal reservoirs by simply introducing common thermal reservoir coupling to both qubits. We find that it is advantageous for the common reservoir to be cold, and that there is a maximal temperature of this reservoir up to which entanglement is enhanced. When the two qubits in the absence of the common reservoir thermalise to a definite temperature -- either because the two independent reservoirs are at thermal equilibrium or because an effective temperature can be assigned to the qubits in the steady state -- then this upper bound is simply equal to the thermalisation temperature. In all cases where entanglement is enhanced, the enhancement is associated with heat absorption by the common reservoir which is thus effectively cooling the system. In the equilibrium case, we observe that with the common reservoir present, entanglement can be generated for larger temperatures of the individual reservoirs than otherwise possible. We have proposed and analysed an implementation of our scheme using superconducting qubits and have seen that even in the presence of additional dephasing and for experimentally accessible parameter settings, a pronounced improvement of steady-state entanglement is possible, and a signficant amount of entanglement can be generated.

\section*{Acknowledgments}

We acknowledge helpful discussions with G. Haack on implementations in superconducting systems. In this work Z.X.M. and Y.J.X. are supported by National Natural Science Foundation (China) under Grant Nos.~11574178 and 61675115, Shandong Provincial Natural Science Foundation (China) under Grant No.~ZR2016JL005, and
Taishan Scholarship Project of Shandong Province. A.T. and J.B.B. acknowledge funding from the Swiss National Science Foundation starting Grant DIAQ, Grant No.~200021 169002.

\appendix
\setcounter{section}{1}

\section*{Appendix: Sketch of master equation derivation}

In this appendix, we sketch the derivation of the master equation (\ref{mast}) in the presence of a common bath. A detailed derivation can be found in Ref.~\cite{equi7} (see in particular Appendix B).

From Eq. (\ref{H-coupling}), the interaction between the systems and the common reservoir is
\begin{equation}
\label{H-com}
\hat{H}_{SR}^{com} = \sum_{n}\left[\left(g_{A,n}\hat{\sigma}_{+}^{A}+g_{B,n}\hat{\sigma}_{+}^{B}\right)\hat{c}_{n}+
\left(g_{A,n}\hat{\sigma}_{-}^{A}+g_{B,n}\hat{\sigma}_{-}^{B}\right)\hat{c}_{n}^{\dag}\right],
\end{equation}
which can be reformulated in the interacting picture with respect to free Hamiltonians of the system and the common reservoir as
\begin{equation}
\label{Hint-com}
\hat{H}_{SR,I}^{com}(t)=\left[\tau_{12}T_{12}(t)+\tau_{34}T_{34}(t)\right]e^{i\omega_{1}t}
+\left[\tau_{13}T_{13}(t)+\tau_{24}T_{24}(t)\right]e^{i\omega_{2}t}+H.c.,
\end{equation}
where $\tau_{ij}=\left|\lambda_{i}\right\rangle\left\langle\lambda_{j}\right|$ and the noise operators
\begin{eqnarray}
  T_{12}(t) &=& \sin\frac{\theta}{2}A(t)+\cos\frac{\theta}{2}B(t), \nonumber\\
 T_{34}(t) &=& -\sin\frac{\theta}{2}A(t)+\cos\frac{\theta}{2}B(t), \nonumber\\
  T_{13}(t) &=& \cos\frac{\theta}{2}A(t)-\sin\frac{\theta}{2}B(t), \nonumber\\
   T_{24}(t) &=& \cos\frac{\theta}{2}A(t)+\sin\frac{\theta}{2}B(t)
\end{eqnarray}
with $A(t)=\sum_{n}g_{A,n}\hat{c}_{n}e^{-i\omega_{c,n}t}$
and $B(t)=\sum_{n}g_{B,n}\hat{c}_{n}e^{-i\omega_{c,n}t}$.

Under the standard Born-Markov approximation, we obtain the master equation for the systems as
\begin{equation}\label{ME}
\dot{\rho}_{S}=-\int_{0}^{\infty}dt' \mathrm{Tr}_{B}\left[\hat{H}_{SR,I}^{com}(t),\left[\hat{H}_{SR,I}^{com}(t-t'),
\rho_{S}(t)\otimes\rho_{B}\right]\right],
\end{equation}
where $\rho_B$ is the state of the common reservoir (which we will take to be a thermal state) and $\mathrm{Tr}_{B}$ denotes the trace over this reservoir. If we further adopt a rotating-wave approximation, we find
\begin{eqnarray}\label{ME2}
\dot{\rho}_{S}&=&\sum_{(i,j)}\left[\tau_{ij}\rho_{S}\tau_{ji}\int_{0}^{\infty}dt'e^{i(E_{i}-E_{j})t'}\left\langle T_{ij}^{\dag}(-t')
T_{ij}(0)\right\rangle\right.\nonumber\\
&&-\tau_{jj}\rho_{S}\int_{0}^{\infty}dt'e^{-i(E_{i}-E_{j})t'}\left\langle T_{ij}^{\dag}(0)
T_{ij}(-t')\right\rangle \nonumber\\
&&+\tau_{ji}\rho_{S}\tau_{ij}\int_{0}^{\infty}dt'e^{-i(E_{i}-E_{j})t'}\left\langle T_{ij}(-t')
T_{ij}^{\dag}(0)\right\rangle\nonumber\\
&&\left.-\tau_{ii}\rho_{S}\int_{0}^{\infty}dt'e^{i(E_{i}-E_{j})t'}\left\langle T_{ij}(0)
T_{ij}^{\dag}(-t')\right\rangle \right]\nonumber\\
&&+\sum_{(ij,kl)}\left[\tau_{ij}\rho_{S}\tau_{kl}\int_{0}^{\infty}dt'e^{i(E_{i}-E_{j})t'}\left\langle T_{lk}^{\dag}(-t')
T_{ij}(0)\right\rangle\right.\nonumber\\
&&\left.+\tau_{lk}\rho_{S}\tau_{ji}\int_{0}^{\infty}dt'e^{i(E_{i}-E_{j})t'}\left\langle T_{ij}^{\dag}(-t')
T_{lk}(0)\right\rangle\right]+H.c.,
\end{eqnarray}
where the first sum runs over $(i,j)=(1,2),(1,3),(2,3)$, and $(2,4)$, while the second runs over $(ij,kl)=(12,34),(13,42),(31,24)$, and $(43, 12)$. The bath correlation functions, appearing under the integrals, are defined as $\left\langle X(t)Y(t')\right\rangle=\mathrm{Tr}_{B}\left[X(t)Y(t')\rho_{B}\right]$.

The real part of the integrals of the correlation functions determine the dissipation rates entering in the final master equation, while the imaginary parts contribute Lamb-type shifts to the Hamiltonian entering in the master equation. As we argue below, the latter are small and can be neglected, as we do in the main text. First, we give an example of the derivation of a dissipation rate.

Consider the correlation function $\langle T_{12}^{\dag}(-t')T_{12}(0)\rangle$. We have
\begin{eqnarray}\label{eg}
&&\int_{0}^{\infty}e^{i\omega_{1}t'}\left\langle T_{12}^{\dag}(-t')T_{12}(0)\right\rangle dt'\nonumber\\
&=&\sin^{2}(\theta/2)\int_{0}^{\infty}e^{i\omega_{1}t'}\left\langle A^{\dag}(-t')A(0)\right\rangle dt'\nonumber\\
&&+\cos^{2}(\theta/2)\int_{0}^{\infty}e^{i\omega_{1}t'}\left\langle B^{\dag}(-t')B(0)\right\rangle dt'\nonumber\\
&&+\frac{1}{2}\sin\theta \int_{0}^{\infty}e^{i\omega_{1}t'}\left\langle A^{\dag}(-t')B(0)\right\rangle dt'\nonumber\\
&&+\frac{1}{2}\sin\theta \int_{0}^{\infty}e^{i\omega_{1}t'}\left\langle B^{\dag}(-t')A(0)\right\rangle dt'.
\end{eqnarray}
As shown in \cite{equi7}, the real parts of the four integrals on the right-hand side of Eq. (\ref{eg}) can be obtained by using the formula
\begin{equation}\label{ff}
\int_{0}^{\infty}dt'e^{\pm i\omega t'}=\pi\delta(\omega)\pm iP\frac{1}{\omega}
\end{equation}
where $P$ denotes the Cauchy principal value integral. For a reservoir in a thermal state, one obtains
\begin{eqnarray}
\textrm{Re} \left[ \int_{0}^{\infty}e^{i\omega_{1}t'}\left\langle A^{\dag}(-t')A(0)\right\rangle dt' \right] & = & \Gamma_C^{(A)}(\omega_1) \bar{n}(\omega_1) , \nonumber \\
\textrm{Re} \left[ \int_{0}^{\infty}e^{i\omega_{1}t'}\left\langle B^{\dag}(-t')B(0)\right\rangle dt' \right] & = & \Gamma_C^{(B)}(\omega_1) \bar{n}(\omega_1) , \nonumber \\
\textrm{Re} \left[ \int_{0}^{\infty}e^{i\omega_{1}t'}\left\langle A^{\dag}(-t')B(0)\right\rangle dt' \right] & = & \Gamma_{C}^{(AB)}(\omega_1) \bar{n}(\omega_1) , \nonumber \\
\textrm{Re} \left[ \int_{0}^{\infty}e^{i\omega_{1}t'}\left\langle B^{\dag}(-t')A(0)\right\rangle dt' \right] & = & \Gamma_{C}^{(AB)}(\omega_1) \bar{n}(\omega_1)  .
\end{eqnarray}
Here, the rates on the right-hand side fulfil $\Gamma_{C}^{(AB)}(\omega_1) = \sqrt{\Gamma_C^{(A)}(\omega_1) \Gamma_C^{(B)}(\omega_1)}$ and are determined by the reservoir density of states and the system-bath coupling coefficients $g_{A,n}$, $g_{B,n}$.

The imaginary part of the one-sided Fourier transform integrals is related to the real part by a Cauchy principal value integral of the form
\begin{equation}
f_I(\omega) =  \int_0^\infty \frac{f_R(\omega')}{\omega - \omega'} d\omega' ,
\end{equation}
where $f_R$ and $f_I$ stand for the real and imaginary parts of integrals over the correlation functions as in the expressions above. In the master equation, they enter in the Hamiltonian part, as shifts of the system energy levels (i.e.~Lamb shifts). Provided that the system-bath couplings (and hence the dissipation rates $\gamma_k$) are small, these imaginary parts will be small relative to the system energy splittings, and so can be neglected.

\section*{References}
\providecommand{\newblock}{}


\begin{thebibliography}{10}
\expandafter\ifx\csname url\endcsname\relax
  \def\url#1{{\tt #1}}\fi
\expandafter\ifx\csname urlprefix\endcsname\relax\def\urlprefix{URL }\fi
\providecommand{\eprint}[2][]{\url{#2}}

\bibitem{ent}
Nielsen M~A and Chuang I~L 2007 {\em Quantum Computation and Quantum
  Information\/} (Cambridge, UK: Cambridge University Press)

\bibitem{giovanetti2011}
Giovannetti V, Lloyd S and Maccone L 2011 {\em Nat. Photonics\/} {\bf 5} 222

\bibitem{pur1}
Bennett C~H, Brassard G, Popescu S, Schumacher B, Smolin J~A and Wootters W~K
  1996 {\em Phys. Rev. Lett.\/} {\bf 76} 722

\bibitem{pur2}
Pan J~W, Gasparoni S, Ursin R, Weihs G and Zeilinger A 2003 {\em Nature\/} {\bf
  423} 417

\bibitem{pur3}
Kwiat P~G, Barrazalopez S, Stefanov A and Gisin N 2001 {\em Nature\/} {\bf 409}
  1014

\bibitem{error-corr1}
Shor P~W 1995 {\em Phys. Rev. A\/} {\bf 52} R2493

\bibitem{error-corr2}
Steane A~M 1996 {\em Phys. Rev. Lett.\/} {\bf 77} 793

\bibitem{DD1}
Mukhtar M, Saw T~B, Soh W~T and Gong J 2010 {\em Phys. Rev. A\/} {\bf 81}
  012331

\bibitem{DD2}
Wang Z~Y and Liu R~B 2011 {\em Phys. Rev. A\/} {\bf 83} 022306

\bibitem{DD3}
Lo~Franco R, D'Arrigo A, Falci G, Compagno G and Paladino E 2014 {\em Phys.
  Rev. B\/} {\bf 90} 054304

\bibitem{Zeno1}
Maniscalco S, Francica F, Zaffino R~L, Lo~Gullo N and Plastina F 2008 {\em
  Phys. Rev. Lett.\/} {\bf 100} 090503

\bibitem{Zeno2}
An N~B, Kim J and Kim K 2010 {\em Phys. Rev. A\/} {\bf 82} 032316

\bibitem{weak1}
Sun Q, Al-Amri M, Davidovich L and Zubairy M~S 2010 {\em Phys. Rev. A\/} {\bf
  82} 052323

\bibitem{weak2}
Kim Y~S, Lee J~C, Kwon O and Kim Y~H 2011 {\em Nat. Phys.\/} {\bf 8} 117

\bibitem{weak3}
Man Z~X, Xia Y~J and An N~B 2012 {\em Phys. Rev. A\/} {\bf 86} 012325

\bibitem{weak4}
Man Z~X, Xia Y~J and An N~B 2012 {\em Phys. Rev. A\/} {\bf 86} 052322

\bibitem{diss-pre1}
Plenio M~B, Huelga S~F, Beige A and Knight P~L 1999 {\em Phys. Rev. A\/} {\bf
  59} 2468

\bibitem{diss-pre2}
Bellomo B, Lo~Franco R, Maniscalco S and Compagno G 2008 {\em Phys. Rev. A\/}
  {\bf 78} 060302

\bibitem{diss-pre3}
Diehl S, Micheli A, Kantian A, Kraus B, Buchler H~P and Zoller P 2008 {\em Nat.
  Phys.\/} {\bf 4} 878

\bibitem{diss-pre4}
Verstraete F, Wolf M~M and Cirac J~I 2009 {\em Nat. Phys.\/} {\bf 5} 633

\bibitem{diss-pre5}
Kraus B, B\"uchler H~P, Diehl S, Kantian A, Micheli A and Zoller P 2008 {\em
  Phys. Rev. A\/} {\bf 78} 042307

\bibitem{diss-pre6}
Ticozzi F and Viola L 2014 {\em Quantum Inf. Comput.\/} {\bf 14} 265

\bibitem{phys-sys7}
Plenio M~B and Huelga S~F 2002 {\em Phys. Rev. Lett.\/} {\bf 88} 197901

\bibitem{phys-sys1}
Schneider S and Milburn G~J 2002 {\em Phys. Rev. A\/} {\bf 65} 042107

\bibitem{phys-sys2}
Kastoryano M~J, Reiter F and S\o{}rensen A~S 2011 {\em Phys. Rev. Lett.\/} {\bf
  106} 090502

\bibitem{phys-sys3}
Reiter F, Tornberg L, Johansson G and S\o{}rensen A~S 2013 {\em Phys. Rev. A\/}
  {\bf 88} 032317

\bibitem{phys-sys4}
Schuetz M~J~A, Kessler E~M, Vandersypen L~M~K, Cirac J~I and Giedke G 2013 {\em
  Phys. Rev. Lett.\/} {\bf 111} 246802

\bibitem{phys-sys5}
Cai J~M, Popescu S and Briegel H~J {\em Phys. Rev. E\/} {\bf 82} 021921

\bibitem{phys-sys6}
Walter S, Budich J~C, Eisert J and Trauzettel B {\em Phys. Rev. B\/} {\bf 88}
  035441

\bibitem{atom}
Krauter H, Muschik C~A, Jensen K, Wasilewski W, Petersen J~M, Cirac J~I and
  Polzik E~S 2011 {\em Phys. Rev. Lett.\/} {\bf 107}(8) 080503

\bibitem{ion1}
Barreiro J~T, Muller M, Schindler P, Nigg D, Monz T, Chwalla M, Hennrich M,
  Roos C, Zoller P and Blatt R 2011 {\em Nature\/} {\bf 470} 486

\bibitem{ion2}
Lin Y~J, Gaebler J~P, Reiter F, Tan T~R, Bowler R~S, Sorensen A~S, Leibfried D
  and Wineland D~J 2013 {\em Nature\/} {\bf 504} 415

\bibitem{super}
Shankar S, Hatridge M, Leghtas Z, Sliwa K, Narla A, Vool U, Girvin S~M, Frunzio
  L, Mirrahimi M and Devoret M~H 2013 {\em Nature\/} {\bf 504} 419

\bibitem{equi1}
Arnesen M~C, Bose S and Vedral V 2001 {\em Phys. Rev. Lett.\/} {\bf 87} 017901

\bibitem{equi2}
Wang X 2001 {\em Phys. Rev. A\/} {\bf 64} 012313

\bibitem{equi3}
Wang X 2001 {\em Phys. Lett. A\/} {\bf 281} 101

\bibitem{equi4}
Gunlycke D, Kendon V~M, Vedral V and Bose S 2001 {\em Phys. Rev. A\/} {\bf 64}
  042302

\bibitem{equi5}
Lagmago~Kamta G and Starace A~F 2002 {\em Phys. Rev. Lett.\/} {\bf 88} 107901

\bibitem{equi6}
Canosa N and Rossignoli R 2006 {\em Phys. Rev. A\/} {\bf 73} 022347

\bibitem{equi7}
Liao J~Q, Huang J~F and Kuang L~M 2011 {\em Phys. Rev. A\/} {\bf 83} 052110

\bibitem{nonequi1}
Eisler V and Zimbor\'as Z 2005 {\em Phys. Rev. A\/} {\bf 71} 042318

\bibitem{nonequi2}
Quiroga L, Rodr\'{\i}guez F~J, Ram\'{\i}rez M~E and Par\'{\i}s R 2007 {\em
  Phys. Rev. A\/} {\bf 75} 032308

\bibitem{nonequi3}
Sinaysky I, Petruccione F and Burgarth D 2008 {\em Phys. Rev. A\/} {\bf 78}
  062301

\bibitem{nonequi4}
Huang X~L, Guo J~L and Yi X~X 2009 {\em Phys. Rev. A\/} {\bf 80} 054301

\bibitem{nonequi5}
Pumulo N, Sinayskiy I and Petruccione F 2011 {\em Phys. Lett. A\/} {\bf 375}
  3157

\bibitem{Brunner2014}
Brunner N, Huber M, Linden N, Popescu S, Silva R and Skrzypczyk P 2014 {\em
  Phys. Rev. E\/} {\bf 89} 032115

\bibitem{BraskPRE2015}
Brask J~B and Brunner N 2015 {\em Phys. Rev. E\/} {\bf 92} 062101

\bibitem{Brask2015}
Brask J~B, Haack G, Brunner N and Huber M 2015 {\em New J. Phys.\/} {\bf 17}
  113029

\bibitem{Tavakoli2017}
Tavakoli A, Haack G, Huber M, Brunner N and Brask J~B 2017 {\em arXiv
  e-print\/}  1708.01428

\bibitem{MSKim}
Kim M~S, Lee J, Ahn D and Knight P~L 2002 {\em Phys. Rev. A\/} {\bf 65} 040101

\bibitem{nonint1}
Braun D 2002 {\em Phys. Rev. Lett.\/} {\bf 89} 277901

\bibitem{nonint2}
Benatti F, Floreanini R and Piani M 2003 {\em Phys. Rev. Lett.\/} {\bf 91}
  070402

\bibitem{BB1}
Bellomo B and Antezza M 2013 {\em EPL\/} {\bf 104} 10006

\bibitem{BB2}
Bellomo B and Antezza M 2015 {\em Phys. Rev. A\/} {\bf 91} 042124

\bibitem{BB3}
Bellomo B and Antezza M 2013 {\em New J. Phys.\/} {\bf 15} 113052

\bibitem{Hofer2017}
Hofer P~P, Perarnau-Llobet M, Miranda L~D~M, Haack G, Silva R, Brask J~B and
  Brunner N 2017 {\em New Journal of Physics\/} {\bf 19} 123037

\bibitem{def-curr1}
Breuer H~P and Petruccione F 2002 {\em The Theory of Open Quantum Systems\/}
  (Oxford: Oxford University Press)

\bibitem{Rivas2010}
Rivas A, Plato A~D~K, Huelga S~F and Plenio M~B 2010 {\em New J. Phys.\/} {\bf
  12} 113032

\bibitem{schaller2015}
Schaller G 2015 {\em Lecture notes: Non-Equilibrium Master Equations\/}
  (Technische Universit\"at Berlin)

\bibitem{NC}
Higgins K~D~B, Benjamin S~C, Stace T, Milburn G, Lovett B and Gauger E 2014
  {\em Nat. Commun.\/} {\bf 5} 4705

\bibitem{Kolodynski2017}
Ko\l{}ody\'{n}ski J, Brask J~B, Perarnau-Llobet M and Bylicka B 2017 {\em arXiv
  e-print\/}  1704.08702

\bibitem{con}
Wootters W~K 1998 {\em Phys. Rev. Lett.\/} {\bf 80} 2245

\bibitem{xtype}
Yu T and Eberle J~H 2007 {\em Quantum Inf. Comput.\/} {\bf 7} 459

\bibitem{Wichterich2007}
Wichterich H, Henrich M~J, Breuer H~P, Gemmer J and Michel M 2007 {\em Phys.
  Rev. E\/} {\bf 76} 031115

\bibitem{def-curr2}
Nieuwenhuizen T~M and Allahverdyan A~E 2002 {\em Phys. Rev. E\/} {\bf 66}
  036102

\bibitem{Teff1}
Quan H~T, Zhang P and Sun C~P 2005 {\em Phys. Rev. E\/} {\bf 72} 056110

\bibitem{Teff2}
Quan H~T, Liu Y~x, Sun C~P and Nori F 2007 {\em Phys. Rev. E\/} {\bf 76} 031105

\bibitem{Teff3}
Quan H~T, Wang Y~D, Liu Y~x, Sun C~P and Nori F 2006 {\em Phys. Rev. Lett.\/}
  {\bf 97} 180402

\bibitem{Cottet2017}
Cottet N, Jezouin S, Bretheau L, Campagne-Ibarcq P, Ficheux Q, Anders J,
  Auff\`{e}ves A, Azouit R, Rouchon P and Huard B 2017 {\em PNAS\/} {\bf 114}
  7561

\bibitem{Koski2013}
Koski J~V, Sagawa T, Saira O~P, Yoon Y, Kutvonen A, Solinas P, Mottonen M,
  Ala-Nissila T and Pekola J~P 2013 {\em Nat. Phys.\/} {\bf 9} 644

\bibitem{Koski2014}
Koski J~V, Maisi V~F, Pekola J~P and Averin D~V 2014 {\em PNAS\/} {\bf 111}
  13786

\bibitem{Pekola2015}
Pekola J~P 2015 {\em Nat. Phys.\/} {\bf 11} 118

\bibitem{Chen2012}
Chen Y~X and Li S~W 2012 {\em Europhys. Lett.\/} {\bf 97} 40003

\bibitem{Hofer2016a}
Hofer P~P, Souquet J~R and Clerk A~A 2016 {\em Phys. Rev. B\/} {\bf 93} 041418

\bibitem{Hofer2016b}
Hofer P~P, Perarnau-Llobet M, Brask J~B, Silva R, Huber M and Brunner N 2016
  {\em Phys. Rev. B\/} {\bf 94} 235420

\bibitem{Manucharyan2012}
Manucharyan V~E 2012 {\em Superinductance\/} Ph.D. thesis Yale University

\bibitem{Majer2007}
Majer J, Chow J~M, Gambetta J~M, Koch J, Johnson B~R, Schreier J~A, Frunzio L,
  Schuster D~I, Houck A~A, Wallraff A, Blais A, Devoret M~H, Girvin S~M and
  Schoelkopf R~J 2007 {\em Nature\/} {\bf 449} 443

\bibitem{Sillanpaa2007}
Sillanp{\"a}{\"a} M~A, Park J~I and Simmonds R~W 2007 {\em Nature\/} {\bf 449}
  438

\bibitem{DiCarlo2009}
DiCarlo L, Chow J~M, Gambetta J~M, Bishop L~S, Johnson B~R, Schuster D~I, Majer
  J, Blais A, Frunzio L, Girvin S~M and Schoelkopf R~J 2009 {\em Nature\/} {\bf
  460} 240

\bibitem{Chen2014}
Chen Y and \textit{et al} 2014 {\em Phys. Rev. Lett.\/} {\bf 113} 220502

\bibitem{Filipp2011}
Filipp S, G\"oppl M, Fink J~M, Baur M, Bianchetti R, Steffen L and Wallraff A
  2011 {\em Phys. Rev. A\/} {\bf 83} 063827

\bibitem{Kim2015}
Kim M~D 2015 {\em Quantum Inf. Proce.\/} {\bf 14} 3677

\bibitem{OMalley2016}
O'Malley P~J~J 2016 {\em Superconducting Qubits: Dephasing and Quantum
  Chemistry\/} Ph.D. thesis University of California, Santa Barbara

\bibitem{Kou2017}
Kou A, Smith W~C, Vool U, Pop I~M, Sliwa K~M, Hatridge M~H, Frunzio L and
  Devoret M~H 2017 {\em arXiv e-print\/}  1705.05712

\end{thebibliography}
\end{document}